\documentclass[11pt,a4paper]{article}
\usepackage{jheppub}
\usepackage{graphics} 
\usepackage{epsfig}
\usepackage{amsmath,amssymb,amsfonts,textcomp}
\usepackage{hyperref}
\usepackage{empheq}
\usepackage{mathrsfs}
\usepackage{eucal,keyval}
\usepackage{comment}

\newcommand{\eq}[1]{Eq.~\eqref{eq:#1}}

\def\dpdf#1{F_{#1}}
\def\spdf#1{q_{#1}}

\newcommand{\bbc}{}
\newcommand{\nn}{\nonumber}
\newcommand{\df}{\mathrm{d}}
\newcommand{\img}{\mathrm{i}}

\newcommand{\si}{\sigma}

\newcommand{\lqcd}{\Lambda_\mathrm{QCD}}

\title{Double parton correlations and constituent
quark models: a Light Front approach to the valence sector}

\author[a]{M. Rinaldi}
\author[a]{S. Scopetta}
\author[b]{M. Traini}
\author[c]{V. Vento}

\affiliation[a]
{Dipartimento di Fisica e Geologia, Universit\`a degli Studi di Perugia, and
INFN, sezione di Perugia, \\ via A. Pascoli
06100 Perugia, Italy}

\affiliation[b]
{Dipartimento di Fisica, Universit\`a degli studi di Trento, and
INFN - TIFPA, \\
Via Sommarive 14, I - 38123 Povo (Trento), Italy}

\affiliation[c]
{Departament de Fisica Te\`orica, Universitat de Val\`encia
and Institut de Fisica Corpuscular, Consejo Superior de Investigaciones
Cient\'{\i}ficas, \\
46100 Burjassot (Val\`encia), Spain}

\emailAdd{matteo.rinaldi@pg.infn.it}
\emailAdd{sergio.scopetta@pg.infn.it}
\emailAdd{traini@science.unitn.it}
\emailAdd{vicente.vento@uv.es}

\abstract{
An explicit evaluation of the double parton distribution functions
(dPDFs), within a relativistic Light-Front approach to
constituent quark models, is presented. 
dPDFs encode information on the correlations
between two partons inside a target and represent 
the non-perturbative QCD ingredient for the description of
double parton scattering in proton-proton collisions, 
a crucial issue in the search of new Physics at the LHC.  
Valence dPDFs are evaluated at the low scale of the model
and the perturbative scale of the experiments
is reached by means of QCD evolution.
The present results show that the strong
correlation effects present at the scale of the model
are still sizable, in the valence region, at the experimental scale.
At the low values of $x$ presently studied at the LHC the
correlations become less relevant, although they
are still important for the spin-dependent contributions 
to unpolarized proton scattering.}

\begin{document}

\maketitle

\section{Introduction}

Multiple hard partonic collisions occurring 
in a single hadronic scattering, 
the so called multiple parton interactions (MPI),
{have been studied since a long time ago~\cite{Paver:1982yp}}.
MPI are suppressed by a power of 
$\lqcd^2/Q^2$ with respect to single parton
interactions,
$Q$ {being} the partonic center-of-mass energy 
in the collision. Technically, they are therefore
higher twist distributions. Despite of this,
experimental evidence of these processes
has been obtained already several years ago \cite{livio}.
At the LHC, MPI, representing
a background for the search of new Physics, 
are of great relevance.
In recent years therefore 
a strong debate around MPI has arised
(see Refs. \cite{Gaunt:2009re,Diehl:2011yj,Manohar:2012jr}
for comprehensive papers on the subject). 
Several dedicated workshops have been organized, starting from that 
illustrated in Ref. \cite{pg}. 

The subject of this work is related to double parton scattering (DPS). 
It is now understood that DPS 
contributes to same-sign $WW$ and same-sign 
dilepton productions
~\cite{Kulesza:1999zh,Cattaruzza:2005nu,Maina:2009sj,Gaunt:2010pi}.
New signatures of DPS, i.e., double Drell-Yan processes,
have been also identified
~\cite{Kasemets:2012pr}.
DPS represents besides a background for Higgs studies in the channel 
$pp \to WH \to \ell \nu b\bar b$ 
\cite{DelFabbro:1999tf,Hussein:2006xr,Bandurin:2010gn,Berger:2011ep}. 
Evidence of DPS at the LHC has been established~\cite{Aad:2013bjm}.

In their seminal work, the authors of Ref.~\cite{Paver:1982yp}
wrote the DPS cross section in terms
of double parton distribution functions (dPDFs), 
$\dpdf{ij}(x_1,x_2,\vec z_\perp)$, describing the joint probability 
of having two partons with flavors $i,j=q, \bar q,g$, 
longitudinal momentum fractions $x_1,x_2$ and transverse separation 
$\vec z_\perp$ inside a hadron:
\begin{align}  \label{eq:si_old}
  \df \si &= \frac{1}{S} \sum_{i,j,k,l} \int\! \mathbf{\df} \vec z_\perp\, 
\dpdf{ij}(x_1,x_2,\vec z_\perp,\mu) 
\dpdf{kl}(x_3,x_4,\vec z_\perp,\mu)  \nn \\ & \quad \times 
  \hat \si_{ik}(x_1 x_3 \sqrt{s},\mu) \hat \si_{jl}(x_2 x_4 \sqrt{s},\mu)
\,.\end{align}
The partonic cross sections $\hat \si$ refer to the hard,
short-distance processes,
$S$ is a symmetry factor, present if identical particles appear in 
the final state and $\mu$ is the renormalization scale. 
For clarity of presentation, such a scale has been taken to be 
the same for both partons, which
is not the case in actual processes, in general.

In \eq{si_old}, contributions due 
to flavor, spin and color correlations between the two partons,
present in QCD, are neglected,  as well as parton-exchange interference 
contributions~\cite{Diehl:2011yj,Manohar:2012jr,Mekhfi:1985dv,Diehl:2011tt}.
Two main assumptions are usually made, for the dPDFs, in DPS analyses: 
\\
i)
the dependences upon the 
transverse separation and the momentum fractions 
or parton flavors are not correlated:
\begin{equation} 
\label{app2}
  \dpdf{ij}(x_1,x_2,\vec z_\perp,\mu) = \dpdf{ij}(x_1,x_2,\mu) 
T(\vec z_\perp,\mu)
\,;\end{equation}
ii)
a factorized form is chosen also for the dependence upon $x_1,x_2$:
\begin{equation} 
\label{app1}
\begin{split}
  &\dpdf{ij}(x_1,x_2,\mu) \\
   &= \spdf{i}(x_1,\mu) \spdf{j}(x_2,\mu)\, \theta(1-x_1-x_2) (1-x_1-x_2)^n
\,,
\end{split}
\end{equation}
where $q$ is the usual parton distribution function (PDF).  
The expression $\theta(1-x_1-x_2) (1-x_1-x_2)^n$,
where $n>0$ is a parameter to be fixed phenomenologically,
introduces the kinematic constraint $x_1+x_2 \leq 1$. 

dPDFs, describing soft Physics,
are nonperturbative objects. 
The dynamical origin of double parton correlations,
having potential effects in the dPDFs and, in turn, in DPS,
has been discussed 
in semi-inclusive deep inelastic scattering and in hard
exclusive processes
\cite{weiss}.
Positivity bounds have been obtained 
for polarized dPDFs \cite{oggi}.
Being non perturbative,
dPDFs have not been evaluated in QCD. As it happens for the usual
PDFs, they can be at least estimated  
at a low scale, 
$Q_0 \sim \lqcd$, the so called hadronic scale, using quark models.
The results of these calculations
should be then evolved using perturbative QCD (pQCD)
in order to compare them with data taken at a momentum scale
$Q>Q_0$, according to a well established procedure, proposed already
in Refs. \cite{pape,jaro}.
The evolution of dPDFs, 
namely the way they change from  
$Q_0$ to $Q>Q_0$, 
known since a long time ~\cite{Kirschner,Shelest:1982dg}, is
currently systematically studied~\cite{Diehl:2011yj,Manohar:2012jr,
Diehl:2011tt,agg1,agg2,Ceccopieri:2010kg,agg3,agg4,
Manohar:DPS2,arriola,kasemets2,
snigirev_evo}.
The result of these analyses is important to relate 
not only data from different experiments with each other, but also model 
calculations at the hadronic scale to data taken at high energy.  
In this way, the analysis of data involving DPS can be guided.

The first model calculation of dPDFs in the valence region
has been presented in Ref.~\cite{man_bag}, in
a bag model framework for the proton,
at the hadronic scale $Q_0$, without evolution 
to $Q>Q_0$.
In a model where
the valence quarks carry all the momentum, such as the bag model,
{the scale $Q_0$} has to be taken quite low.
If the bag were assumed to be rigid,
as in the so-called cavity approximation
~\cite{Jaffe:1974nj},
the quarks would be independent and none of the relevant
correlations described by dPDFs would be found.
In Ref.~\cite{man_bag}, therefore, a prescription is used to recover
momentum conservation, 
already applied in model calculations of PDFs (see, e.g. 
~\cite{Benesh:1987ie}).
In this way, quark correlations in the bag are found.
The analysis of Ref.~\cite{man_bag} has been retaken in a constituent
quark model (CQM) framework in Ref.~\cite{nostro}.
CQM calculations of parton distributions have been proven
to be able to predict the gross features of PDFs \cite{trvlc,h1,orb},
generalized parton distributions (GPDs) \cite{vlc_pg}
and transverse momentum dependent parton distributions (TMDs) 
\cite{siv_bm,siv_sr,bm_sr}. Similar expectations motivated the analysis
of Ref.~\cite{nostro}.
With respect to the approach of Ref.~\cite{man_bag},
the non relativistic (NR) dynamics includes from the very beginning
correlations into the scheme.
The main result of Refs.~\cite{man_bag} and
\cite{nostro} was that Eq. (\ref{app1}) holds reasonably well, 
but Eq. (\ref{app2}) is strongly violated. 
Actually, problems with Eq. (\ref{app2}) had already been pointed out 
in Refs.~\cite{snig1,Gaunt:2009re,snig2} on a general ground.
In the CQM picture of Ref.~\cite{nostro}, the fact that correlations
are naturally included helped to understand their
dynamical origin, for example that
of the breaking of the approximation Eq. ({\ref{app2}}).
To have predictions in different models is important
of course to understand which, among the features of the results,
are the model dependent ones.

Both the analysis of Refs.~\cite{man_bag} and
\cite{nostro} are somehow incomplete.
The missing items are mainly two.
First of all, for different reasons, they both
lead to the so called
``bad support'' problem. This means that
dPDFs are not vanishing in the forbidden
kinematic region, $x_1+x_2 > 1$. 
In the bag model this is due to the lack of momentum conservation,
i.e., proton states are not momentum eigenstates.
In the CQM calculation this is due to the impossibility
to treat correctly the off-shellness of the interacting
partons. 
Secondly, as already stressed, the results of Refs. ~\cite{man_bag} and
~\cite{nostro} are valid at the hadronic scale
and only in the valence quark region. For these studies to be
directly used in the LHC data analysis, having for the moment high
statistics only for small values of $x$, far from the valence region,
corresponding to high momentum transfer,
additional studies, in particular the pQCD evolution
of the model results, are necessary.

The present work is a step towards an improvement in both the
items listed above. 
This is obtained within a relativistic, 
fully Poincar\'e covariant
Light-Front approach 
(for comprehensive reports see, e.g., \cite{LF,LF0}).
In this framework, successfully applied in Hadronic Physics 
in general (see, e.g., \cite{LF1}) and for the calculation
of parton distributions in particular (see, e.g., \cite{LF2,pasquini,
pasquinipol,pasquevo,pasquinih,marco1,marco2}),
the active particles are on-shell and the ``bad-support'' problem
does not arise. As it will be shown, this fact helps
to recover the symmetries of the results, expected on general
grounds, and to properly evaluate Mellin moments 
of the obtained distributions, for a formally
correct pQCD evolution of the results.
This last procedure is also performed, for the valence sector 
only, for the moment being. This represents an important step towards
a proper treatment of the second issue listed here above. 

The paper is structured as follows.
In the next section, the formalism to evaluate
the dPDFs in a LF CQM is illustrated and the main equations
presented. In the third one, results at the hadronic
scale, obtained within the relativistic
hypercentral CQM \cite{LF2},
already used for the calculation of parton distributions
in Refs. \cite{LF2,pasquini,
pasquinipol,pasquevo,pasquinih,marco1,marco2},
are presented.
In the fourth section, the pQCD evolution
of the calculated dPDFs is illustrated and discussed.
Eventually, conclusions are drawn in the last section.

\section{Double parton distributions in a Light-Front
Constituent quark model}

In this section, the main features of the calculation of the dPDFs
within a LF CQM are described. 
A quantitative analysis of the validity of the approximations Eqs.
(\ref{app2}) and (\ref{app1}) in this framework will be therefore possible.
Among the forms of relativistic dynamics,
the LF is the one with maximum number of kinematical, interaction
independent generators \cite{Dirac}. In particular, LF boosts
are kinematical and boost-invariant states can be
defined, which makes the LF form very convenient when
high momentum transfers are present, such as in DIS processes,
where light-cone Physics naturally arises.
The Poincar\'e covariance obtained in a LF approach
allows to preserve the symmetries of the problem.
For a comprehensive {introduction to LF methods see, 
for example, Refs \cite{LF,LF0}}.

Let us introduce now formally the dPDFs, the subject of our study.
As in the model calculations of Refs. \cite{man_bag,nostro}, 
the quantity of interest 
will be the Fourier- transformed dPDF in momentum space,   
$\dpdf{ij}^{\lambda_1,\lambda_2}(x_1,x_2,{\vec k}_\perp)$:

\begin{eqnarray} 
\label{ft}
\dpdf{ij}^{\lambda_1,\lambda_2}(x_1,x_2,{\vec k}_\perp) 
= \int\! d \vec z_\perp \, e^{\img \vec z_\perp \cdot \vec k_\perp} 
\dpdf{ij}^{\lambda_1,\lambda_2}(x_1,x_2,{\vec z}_\perp)~,
\end{eqnarray}

where the coordinate-space expression
$\dpdf{ij}^{\lambda_1,\lambda_2}(x_1,x_2,{\vec z}_\perp)$
reads, in terms of 
Light-Cone (LC) quantized fields $q_i$ for a quark of flavor $i$,
helicity $\lambda_i$ 
and LC normalized proton states, for an unpolarized proton, as follows
(see, e.g., \cite{Manohar:2012jr})

\begin{eqnarray}
\label{vera} 
\dpdf{ij}^{\lambda_1,\lambda_2}(x_1,x_2,{\vec k}_\perp) & = &
(-8 \pi P^{+}) \dfrac{1}{2} \underset{\lambda}\sum \int\! d 
\vec z_\perp \, e^{\img \vec z_\perp
\cdot \vec k_\perp} 
\\
\nonumber
& \times &
\int \left [ \underset{l}{\overset{3}\prod}  \dfrac{
d z_l^-}{4 \pi} \right]
e^{i x_1 P^+ z_1^-/2}e^{i x_2 P^+ z_2^-/2}e^{-i x_1 P^+ z_3^-/2}
\\
\nonumber
& \times &
\langle \lambda, \vec P = \vec 0  \big|
\hat{\mathcal{O}}_i^1 \left( z_1^-\dfrac{\bar n}{2},z_3^- \dfrac{\bar n}{2}+
\vec z_\perp \right)
\hat{\mathcal O}_j^2 \left( z_2^-\dfrac{\bar n}{2}+
\vec z_\perp,0 \right)
\big|  \vec P = \vec 0,\lambda \rangle~,
\end{eqnarray}

where, for generic 4-vectors $z$ and $z'$, the operator

\begin{eqnarray}
\hat{\mathcal O}_i^k(z,z') = \bar q_i(z) \hat{O}(\lambda_k) q_i(z')
\end{eqnarray}

has been defined, with

\begin{eqnarray}
\hat O(\lambda_k) =
\dfrac{\bar n   \hskip-0.2cm /}{2}    
\dfrac{1 + \lambda_k \gamma_5}{2}~.
\end{eqnarray}
Besides, the light-like four vector,
$\bar n  = (1,0,0,-1)$, and 
the rest frame state of the nucleon with helicity
$\lambda$,
$\big| \vec P = \vec 0,\lambda \rangle$,
have been introduced. 
In Eq. (\ref{vera})
color-correlated and interference dPDFs,
Sudakov suppressed at high energies,
have not been considered
\cite{Mekhfi:1988kj,Manohar:2012jr}.

Here and in the following,
the ``$\pm$'' components
of a four-vector $b$ are defined according to $b^\pm = b_0 \pm b_z$ and 
$x_i = \dfrac{k_i^+}{P^+}$ is the
fraction of the system momentum carried by the parton ``i''. 
Moreover, the notation $\tilde b = (b^+, \vec b_\perp)$ { is} used
for LC vectors.

In order to evaluate the above quantities, use has been made of the 
LC free quark fields \cite{spinor0,spinor}:

\begin{eqnarray}
\label{field}
q_i(\xi) &=& \underset{r}\sum \int \dfrac{d \tilde k}{2 (2 \pi)^3 
\sqrt{k^+}} 
\theta(k^+) e^{-i \xi^- k^+} a_{\tilde k,r}^i~ u_{LF}(\tilde k,r) 
\end{eqnarray}

where the operator
$a_{\tilde k,r}^i$ destroys a quark of flavor $i$, helicity $r$
and 
LC momentum 
$\tilde k$.   

The $u_{LF}(\tilde k,r)$ spinors read \cite{LF0}:

\begin{eqnarray}
u_{LF}(\tilde k,+) &=& \dfrac{m}{\sqrt{k^+}} \left(  
\begin{array}{c}
\dfrac{k^+}{m}\\
\\
\dfrac{k_x+ik_y}{m}\\
\\
1\\
\\
0
\end{array}
 \right);~~   u_{LF}(\tilde k,-) =  
{m \over \sqrt{k^+}}
\left(  
\begin{array}{c}
0\\
\\
1\\
\\
-\dfrac{k_x-ik_y}{m}
\\
\\
\dfrac{k^+}{m}
\end{array}
 \right)~.
\end{eqnarray}


Now one has to establish a direct link between the above 
expressions for the dPDF 
and the proton wave function corresponding to a given LF CQM.
To this aim, the procedure proposed in Ref.
\cite{pasquini} for the model evaluation of GPDs
has been extended to the present case.

Starting from the
general expression Eq. (\ref{vera}),
by expanding the proton state $|\vec 0, \lambda \rangle$
in its Fock components and retaining
only the first,
valence contribution, $|\vec 0, \lambda, val \rangle$, 
one obtains, 
in terms of the LF one-quark states of isospin $\tau_i$,
$|\tilde k_i, \lambda_i^f,
\tau_i \rangle $
\cite{LF0}:

\begin{eqnarray}
\label{f-nstate}
|\vec 0, \lambda \rangle \simeq 
|\vec 0, \lambda^f, val \rangle &=& \underset{\lambda_i^f \tau_i}\sum
\int\left[ \underset{i=1}{\overset{3}\prod}  \dfrac{d x_i}{\sqrt{x_i}}   
\right]
\delta \left( 1- \underset{i=1}{\overset{3}\sum}x_i \right)
\left[
\underset{i=1}{\overset{3}\prod}  \dfrac{d \vec
k_{i\perp}}{2(2\pi)^3}  \right]
\delta \left(  \underset{i=1}{\overset{3}\sum}\vec k_{i\perp} \right)
\\
\nonumber
&\times& 2(2\pi)^3\Psi^{[f]}_{\lambda}(\lbrace x_i, \vec k_{i\perp}
,\lambda_i^f,
\tau_i \rbrace) \underset{i=1}{\overset{3}\prod} |\tilde k_i, \lambda_i^f,
\tau_i \rangle~.
\end{eqnarray}

In terms of the canonical, Instant-Form (IF) one-quark states 
$|\vec k_i, \lambda_i^c,
\tau_i \rangle$,
the same proton state reads instead:

\begin{eqnarray}
|\vec 0, \lambda \rangle \simeq
|\vec 0, \lambda^c, val \rangle 
&=& \underset{\lambda_i^c \tau_i}\sum
\int\left[ \underset{i=1}{\overset{3}\prod}  d \vec k_i 
\right]\delta \left(  \underset{i=1}{\overset{3}\sum}\vec k_{i} \right)
\Psi^{[c]}_{\lambda}(\lbrace  \vec k_{i} ,\lambda_i^c,
\tau_i \rbrace) \underset{i=1}{\overset{3}\prod} |\vec k_i, \lambda_i^c,
\tau_i \rangle~.
\label{I-nstate}
\end{eqnarray}

Here and in the following the short-hand notation $\lbrace \alpha_i 
\rbrace$ is adopted for
$\alpha_1, \alpha_2, \alpha_3$.
The 
orthogonality conditions for the LF and IF states read, respectively:

\begin{eqnarray}
&& \langle \tau, \lambda^f,  \tilde k
| \tilde k', \lambda'^f, \tau' \rangle = 
2(2 \pi)^3  k^+ \delta(k^+-k'^+  )~ \delta^2(\vec k_\perp- 
\vec k'_\perp) \delta_{\tau, \tau'} \delta_{\lambda^f, \lambda'^f}~;
\\
\nonumber
&& \langle \tau, \lambda^c, \vec k | \vec k', \lambda'^c, 
\tau'
\rangle =  \delta^3(\vec k- 
\vec k') \delta_{\tau, \tau'} \delta_{\lambda^c, \lambda'^c}~.
\end{eqnarray}

The above descriptions can be related to each other through the following
fundamental relation
\cite{LF}

\begin{eqnarray}
 | \tilde k, \lambda^f, \tau \rangle = \sqrt{\omega}(2\pi)^{3/2}
\underset{\lambda^c}\sum D^{1/2}_{\lambda^f \lambda^c}(R_{cf}(\vec k)) | \vec
k, \lambda^c, \tau \rangle~,
\label{1pstate}
\end{eqnarray}

expressing
the LF spin state of a particle from its canonical IF one,
where $\omega = \sqrt{m^2+\vec k^2}$ and
the Melosh rotations $D^{1/2}_{\mu \lambda}(R_{cf}(\vec k))$ 
naturally arise.

{  
The above equation relates {\it free} canonical and
light-front states. Actually we are interested in interacting
quarks in a proton and the connection of instant form and
light front states for composite systems is, in this case,
much more complicated. Nevertheless, if
one chooses
a suitable representation of the Poincar\'e operators, 
such as the Bakamjian-Thomas construction \cite{BT},
Eq. (\ref{1pstate}) can be generalized to interacting
states in composite systems \cite{LF2,pasquini,
pasquinipol,pasquevo,pasquinih,marco1,marco2}.
The dynamical framework used in the following consists in a relativistic
mass equation built in accord with
the Bakamjian-Thomas construction. 
For the physical situations relevant to the subject of this paper, 
therefore, one can relate
the valence contributions to the nucleon state in its rest frame, 
in the LF and IF, given by Eqs. (\ref{f-nstate}) and (\ref{I-nstate}) 
respectively,
as follows (see, e.g., Ref. \cite{pasquini}):}

\begin{eqnarray}
|\vec 0, \lambda^f, val \rangle = \sqrt{M_0}(2\pi)^{3/2} |\vec 0,
\lambda^c, val \rangle~,
\label{I-C-nstate}
\end{eqnarray}

where $M_0 = \underset{i}\sum \omega_i$ is the free quarks energy,
in terms of which the free mass, invariant for LF boosts, is defined
as follows:

\begin{eqnarray}
M_0^2 = \sum_{i} { m_i^2 + \vec k_{i\perp}^2 \over x_i}~.
\label{invariant}
\end{eqnarray}

The Melosh operators read, in our notation:

\begin{eqnarray}
D^{1/2}_{\mu
\lambda}(R_{cf}(\vec k_i)) =
\langle \mu |~ \hat{D}_i ~| \lambda \rangle~,
\end{eqnarray}

with

\begin{eqnarray}
\hat{D}_i
= \dfrac{m +x_iM_0 - i \vec 
\sigma_i
\cdot (\hat z \times \vec k_\perp)}{\sqrt{(m +x_i M_0)^2+\vec k_\perp^2}}~.
\end{eqnarray}

Substituting the LF quark states, Eq. (\ref{1pstate}), into 
the proton LF state, Eq. (\ref{f-nstate}), one gets the latter
in terms of IF quark states and Melosh operators.
Then, using Eq. (\ref{I-C-nstate})
for the obtained expression, 
one gets a direct link between the LF
proton wave function (LFWF), $\psi^{[f]}_\lambda$,
appearing in Eq. (\ref{f-nstate}), 
and the
IF one, $\psi^{[c]}_\lambda$,
appearing in Eq. (\ref{I-nstate})
\cite{pasquini}:

\begin{eqnarray}
\label{I-f-wave}
\psi^{[f]}_\lambda (\lbrace x_i, \vec k_{i \perp},
\lambda_i^f,
\tau_i, 
\rbrace) &=& 2(2\pi)^3 \left[\dfrac{\omega_1 \omega_2 \omega_3}{M_0 x_1 x_2
x_3}  \right]^{1/2}
\underset{i=1}{\overset{3} \prod} \left [
\underset{\lambda_i^c}\sum D^{*1/2}_{\lambda_i^c
\lambda_i^f}(R_{cf}(\vec k_i))
\right ]
\nonumber
\\
&\times& \psi^{[c]}_\lambda (\lbrace  \vec k_{i},
\lambda_i^c,
\tau_i
\rbrace)
\\
\nonumber
&=& 2(2\pi)^3 \left[\dfrac{\omega_1 \omega_2 \omega_3}{M_0 x_1 x_2
x_3}  \right]^{1/2}
\Psi(\vec k_1,\vec k_2,\vec k_3; \lbrace \lambda_i^f,\tau_i \rbrace)~.
\end{eqnarray}

{ One should notice that, in principle, 
LFWFs are eigensolutions of the LF hamiltonian
which, in turn, is derived from the QCD Lagrangian.
Our simplified approach, 
as it is apparent from the above equations and as it has
been already stated,
consists in using LFWFs obtained properly boosting
eigenstates of a relativistic effective mass equation,
consistent with the Bakamjian-Thomas construction and 
reproducing some of the QCD symmetries.

In Eq. (\ref{I-f-wave}),} for the sake of convenience, the 
function $\Psi$ has been introduced.
For the model under scrutiny in the following, 
it can be split into a momentum space wave function,
$\psi$, and a spin-orbital-flavor part, as follows:

\begin{eqnarray}
\label{psiint}
\Psi(\vec k_1,\vec k_2,\vec k_2; \lbrace \lambda_i^f, \tau_i \rbrace)
& = &
\psi (\vec k_1, \vec k_2, \vec k_3) 
\underset{i=1}{\overset{3} \prod} \left [
\underset{\lambda_i^c}\sum D^{*1/2}_{\lambda_i^c
\lambda_i^f}(R_{cf}(\vec k_i))
\right ]
\nonumber
\\
&\times &
\Phi(\lambda_1^c,\lambda_2^c,\lambda_3^c, \tau_1, \tau_2, \tau_3)~.
\end{eqnarray}

Now, the wave function Eq. (\ref{I-f-wave}), with
the expression (\ref{psiint}) for $\Psi$, can be used to evaluate 
the valence component of the
LF proton state, Eq. (\ref{f-nstate}). This state is in turn inserted,
together with the quark field expression, 
Eq. (\ref{field}),
in the general definition of the dPDFs, Eq. (\ref{vera}).
Then, using ordinary operator algebra and properly treating the
$\delta$ function on the $x$ variables appearing in Eq.
(\ref{f-nstate}), a straightforward procedure leads to the general
expression for the evaluation of the valence contribution to the dPDF 
Eq. (\ref{ft}) within a given LF CQM.
For quarks of flavor $q_1$ and $q_2$, 
longitudinal momentum fractions $x_1$ and $x_2$ and 
transverse distance $\vec k_\perp$ in momentum space, 
it reads as follows:

\begin{eqnarray}
\label{main}
F_{q_1 q_2}^{\lambda_1,\lambda_2}(x_1, x_2, \vec k_\perp) 
& = & 
3(\sqrt{3})^3 \int
\left[
\underset{i=1}{\overset{3}\prod} d \vec k_i 
\underset{\lambda_i^f \tau_i} {\sum}
\right]
\delta \left(
\underset{i=1}{\overset{3}\sum} \vec k_i 
\right) 
\\
\nonumber
& \times &
\Psi^* \left(\vec k_1 +
\dfrac{\vec k_\perp}{2}, \vec k_2 -
\dfrac{\vec k_\perp}{2},\vec k_3
; \{\lambda_i^f, \tau_i \}
\right) \\
\nonumber
& \times & \widehat{P}_{q_1}(1)\widehat{P}_{q_2}
(2)\widehat { P } _ { \lambda_1 } (1)\widehat { P } _ { \lambda_2 } (2)
\, \Psi \left(\vec k_1 -
\dfrac{\vec k_\perp}{2}, \vec k_2 +
\dfrac{\vec k_\perp}{2}, \vec k_3 
; \{\lambda_i^f, \tau_i \}
\right)
\\
\nonumber
& \times & \delta \left(x_1 
-\dfrac{k_1^+}{P^+}
\right) \delta \left(x_2 -\dfrac{k_2^+}{P^+}
\right)~.
\end{eqnarray}
In order to estimate the dynamical correlations between two
unpolarized or longitudinally polarized quarks,  the  flavor
and spin projectors, acting on the state of the $i$ particle, 
have been introduced in the above equation. They read:

\begin{eqnarray}
\label{proj}
\hat P_{u(d)}(i) & = & \dfrac{ 1 \pm \tau_3(i)}{2}~, \\
\nonumber
\hat P_{\lambda_k}(i) & = & \dfrac{ 1 + \lambda_k \sigma_3(i)}{2}~.
\end{eqnarray}

Eq. (\ref{main}) is the general expression needed to evaluate the
(spin-dependent) dPDFs in a Light-Front CQM.
In particular, 
if $q_1 = q_2 = u$, assuming an SU(6) symmetry for the canonical
proton wave function, as it will be done in this paper, one gets:





\begin{eqnarray}
\label{uuspin}
 u_{\uparrow(\downarrow)}u_{\uparrow(\downarrow)} (x_1,x_2, k_\perp) &=&
 2(\sqrt{3})^3 \int  
 d \vec k_{1\perp} d \vec k_{2\perp}  ~
\dfrac{E_1 E_2 E_3}{k_1^+ x_1 x_2 
(1-x_1-x_2) j} 
 \\
 \nonumber
 &\times&  
\langle \tilde P_1^{\uparrow(\downarrow)}  \rangle 
 \langle \tilde P_2^{\uparrow(\downarrow)}  \rangle \,
 \psi^* 
\left(\vec k_1 +
\dfrac{\vec k_\perp}{2}, \vec k_2 -
\dfrac{\vec k_\perp}{2}, -\vec k_1 -\vec k_2 \right) 
 \\
 \nonumber
 &\times&  
\psi \left(\vec k_1 -
\dfrac{\vec k_\perp}{2}, \vec k_2 +
\dfrac{\vec k_\perp}{2}, -\vec k_1 -\vec k_2 \right).
\end{eqnarray}
The following relations hold among the 
quantities appearing in this equation and the integration variables:

\begin{eqnarray}
 & k_1^+ = \sqrt{x_1 \left\lbrace m^2 \left[1+\dfrac{x_1}{x_2}+
\dfrac{x_1}{1-x_1-x_2} \right] +k_{1 \perp}^2 + \dfrac{x_1}{x_2}
k_{2\perp}^2
+\dfrac{x_1}{1-x_1-x_2}  k_{3\perp}^2\right\rbrace }~; &
 \\
 \nonumber
 \\
 \nonumber
&
k_2^+ = \dfrac{x_2}{x_1} k_1^+, ~~ k_3^+ = \dfrac{1-x_1-x_2}{x_1} k_1^+
,~~~k_{iz} = -\dfrac{m^2+k_{i\perp}^2}{2 k_i^+} +\dfrac{k_i^+}{2}~;&
\\
\nonumber
&E_i = \sqrt{m^2+k_{iz}^2+\vec k_{i\perp}^2}~;&
\\
\nonumber
&j = \left|
\dfrac{m^2+ k_{1 \perp}^2}{2 k_1^{+2}} +\dfrac{m^2+ k_{2 \perp}^2}{2
\dfrac{x_2}{x_1} k_1^{+2}}+\dfrac{m^2+ k_{3 \perp}^2}{2
\dfrac{1-x_1-x_2}{x_1} k_1^{+2}} + \dfrac{1}{2 x_1} \right|~.&
\end{eqnarray}
Besides, the spin structure of Eq. (\ref{uuspin}) is described by the 
coupling of the Melosh rotation with the spin projection operators 
defined in Eq. (\ref{proj}), as follows:

\begin{eqnarray}
&&\tilde{P}^{\uparrow(\downarrow)}_i=\hat D_i 
\hat P_{\uparrow(\downarrow)}(i)
\hat D_i^\dagger~, 
\end{eqnarray}
so that, using the canonical
spin-isospin state corresponding to the SU(6) symmetry,
the matrix elements appearing in Eq. (\ref{uuspin}) read:

\begin{eqnarray}
 \langle  \tilde P^{\uparrow(\downarrow)}_i \rangle &=& 
\left \langle 
\dfrac{\chi_{MS}\phi_{MS}+\chi_{MA}\phi_{MA}}{\sqrt{2}}
\left| \tilde
P^{\uparrow(\downarrow)}_i \right| 
\dfrac{\chi_{MS}\phi_{MS}+\chi_{MA}\phi_{MA}}{
\sqrt{2}}   \right \rangle~.
\end{eqnarray}

As one can see in Eq. (\ref{main}),  the dPDFs evaluated by using
the Light-Front treatment depend on delta functions, defining the
longitudinal
momentum fractions carried by the quarks inside the proton, which are function
of
$P^+$, the plus component of the proton momentum, which, in the LF intrinsic
frame, reads:

\begin{eqnarray}
P^+= \overset{3}{ \underset{i}\sum} k_i^+ =
\overset{3}{ \underset{i}\sum}\omega_i = M_0~, 
\end{eqnarray}
where $M_0$ is the free quarks total energy, already defined
in Eq. (\ref{invariant}).
Thanks to
this relation, the
correct support is obtained, i.e.,  
dPDFs are vanishing in the forbidden
kinematic region, $x_1+x_2 > 1$. This result 
is natural in the LF approach (see, e.g., 
Refs. \cite{pasquini,pasquinipol,pasquevo,pasquinih,marco1,marco2}). 

In particular, in the present investigation, the following combinations of the
spin components of the
expression Eq. (\ref{uuspin}) will be described:

\begin{eqnarray}
 \label{unp}
uu(x_1,x_2, k_\perp) = \underset{i,j = \uparrow,\downarrow}\sum
u_iu_j(x_1,x_2, k_\perp),
\end{eqnarray}
i.e., the dPDF describing two unpolarized quarks, and

\begin{eqnarray}
 \label{pol}
\Delta u \Delta u(x_1,x_2, k_\perp) = \underset{i=
\uparrow,\downarrow}\sum
u_iu_i(x_1,x_2, k_\perp)-\underset{i\neq j =
\uparrow,\downarrow}\sum
u_iu_j(x_1,x_2, k_\perp)~,
\end{eqnarray}

i.e., the one in the  case of two polarized quarks. 
These are  the only
distributions contributing to the total cross section in
events involving  unpolarized
protons (see, e.g., Ref. \cite{Diehl:2011yj}).

\section{Results at the hadronic scale}

In order to calculate the dPDFs,
a wave function within
a proper CQM is needed. In this paper, as an example, 
use will be made of the 
relativistic hyper-central CQM, firstly introduced in Ref. \cite{LF2},
providing a reasonable description of the light baryon spectrum.
This model has been systematically applied to the evaluation 
of parton distributions in Refs.
\cite{LF2,pasquini,pasquinipol,pasquevo,pasquinih,marco1,marco2}.
The present analysis is exploratory since there are no data
available for the observables under scrutiny. This model gives us
sufficient guarantee to be able to grasp the most relevant features of dPDFs.
Together with the LF treatment, it provides a scheme that reproduces
the symmetries which are relevant in the problem.
Details on the construction of the model and on the fixing of its
parameters to low energy properties of the proton can be found
in Refs. \cite{LF2,marco1,marco2}. 
For the discussion here it is enough to know that
the total proton state is given by a momentum space wave function
and a spin-isospin part, dictated
by the SU(6) symmetry.
In the model under scrutiny here the momentum space
wave function, $\psi_{00}$,
does not depend on the angular variables but only 
 on the hyper-radius 
$k_\xi$, being a solution of
the Mass equation

\begin{eqnarray}
 (M_0 + V) \psi_{00}(k_\xi) = \left( \overset{3}{\underset{i}\sum } 
\sqrt{\vec k_i^2 + m^2} -  \dfrac{\tau}{\xi} +
 \kappa \xi \right) \psi_{00} (k_\xi) = M \psi_{00}(k_\xi)~.
 \label{mass}
\end{eqnarray}

$\xi$ is the variable conjugated to $k_\xi$ and the  
parameters of the potential are 
fixed in order to reproduce the essential features of the Nucleonic
spectrum. Their values are 
for the relativistic (for the NR reduction of the mass operator Eq.
(\ref{mass}), see Ref. \cite{giannin}) case: $\tau =
3.30\, (4.59)$ and $\kappa = 
1.80\, (1.61)$ fm$^{-2}$.
The solution of this Mass equation is described, e.g., in Ref. \cite{LF2}.
It is instructive to notice that the intrinsic momentum space wave 
function is {simply} given by $\psi(\vec k_1,
\vec k_2)= \psi(k_\xi)$, where here $k_\xi$ is the hyperradius 
in momentum space,
\begin{eqnarray}
k_\xi = \sqrt{2(\vec k_1^2 + \vec k_2^2+ \vec k_1 \cdot \vec k_2 )}~,
\end{eqnarray}
so that the wave function depends on the product $ \vec k_1
\cdot \vec k_2$. This fact, as it happens in the model
calculations of Ref. \cite{nostro}, 
is responsible for the presence of spin-independent
correlations between the two quarks in the CQM calculation.
In the spin-dependent case, the presence of the Melosh rotations
represents an additional source of (spin) correlations.

\begin{figure}[t]
\begin{minipage}[t] {70 mm}
\vspace{7.0cm}
\includegraphics{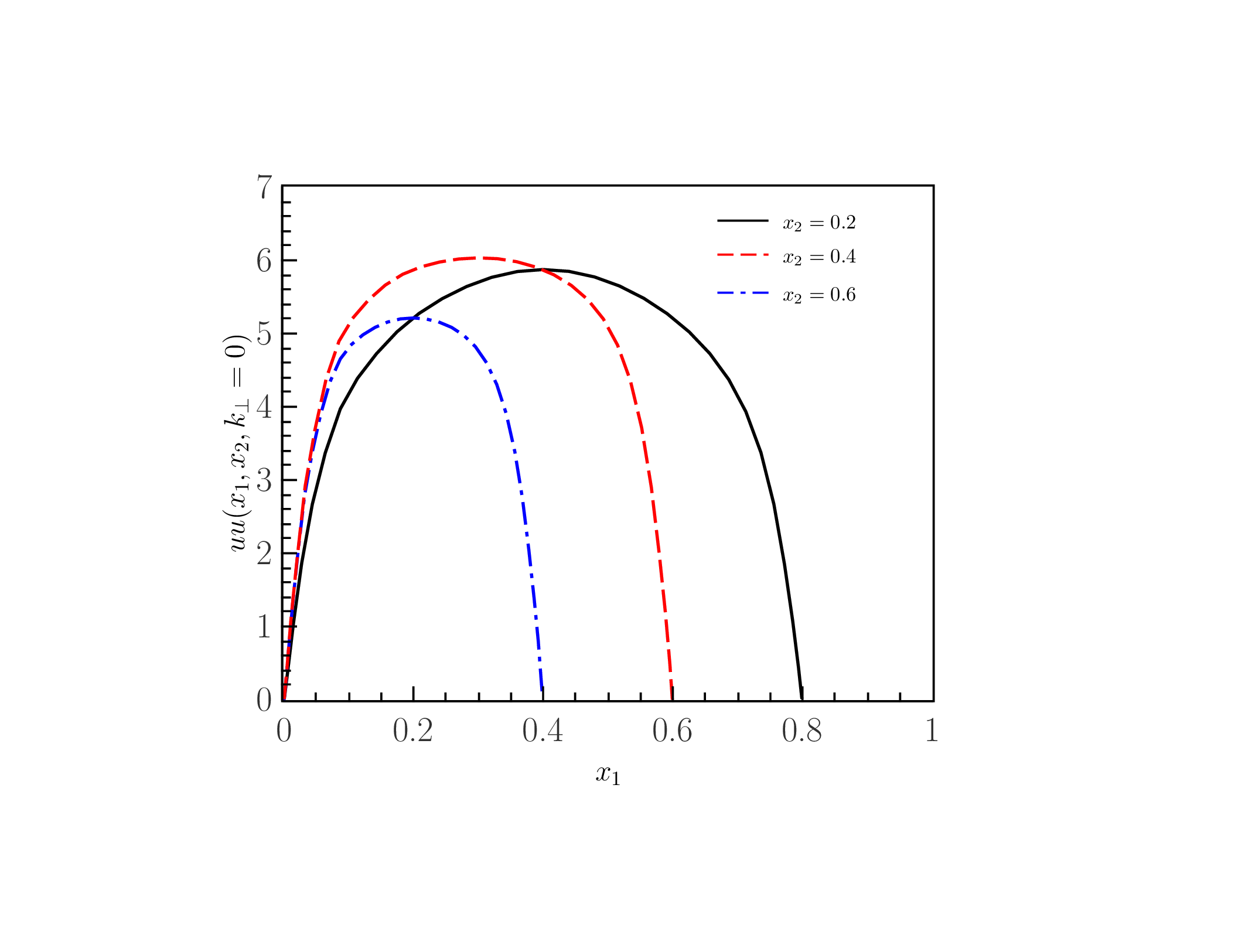}
\caption{ \footnotesize The unpolarized dPDF, Eq. (\ref{unp}), 
as a function of $x_1$ and for
three values of $x_2$, at $k_\perp =0$.}
\end{minipage}
\hspace{\fill}
\begin{minipage}[t] {70 mm}
\vspace{7.0cm}
\includegraphics{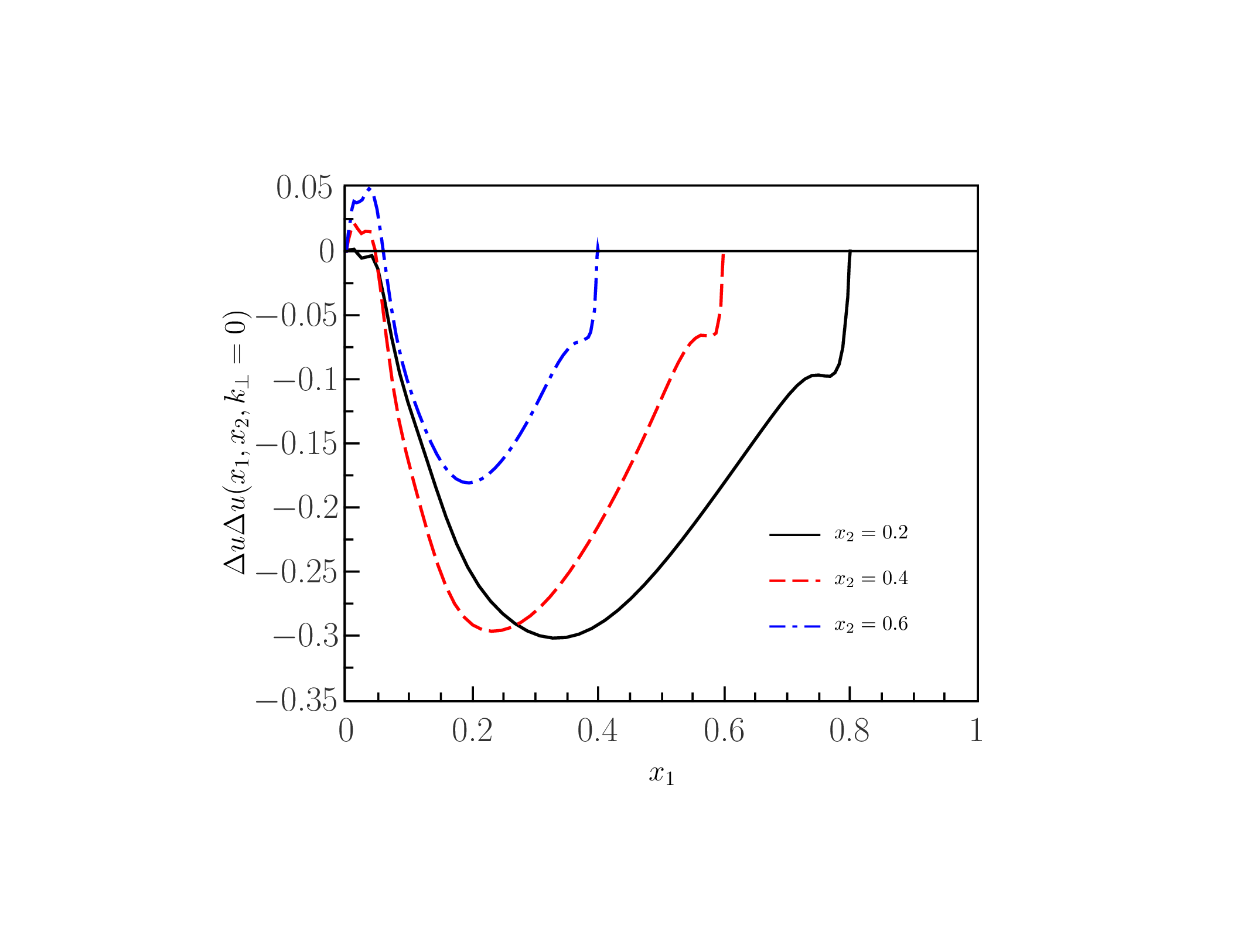}
\caption{ \footnotesize 
The same as in Fig. 1, but for the spin-dependent 
dPDF Eq. (\ref{pol}).}
\end{minipage}
\end{figure}

\begin{figure}[t]
\begin{minipage}[t] {70 mm}
\vspace{7.0cm}
\includegraphics{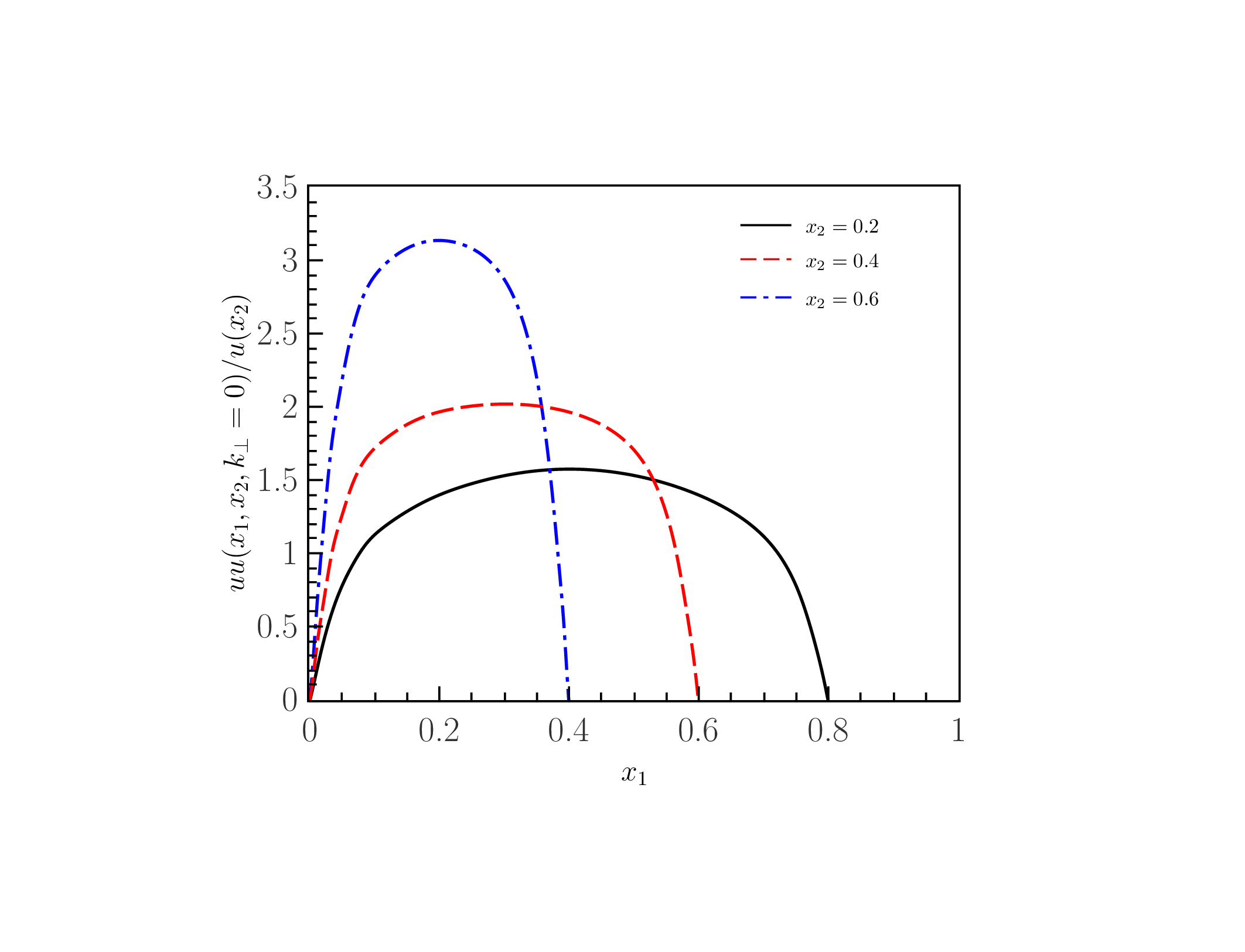}
\caption{\footnotesize The ratio $r_1$, Eq. (\ref{r1}), at $k_\perp =0$,
evaluated for 
three values of $x_2$,
as a function of $x_1$.}
\end{minipage}
\hspace{\fill}
\begin{minipage}[t] {70 mm}
\vspace{7.0cm}
\includegraphics{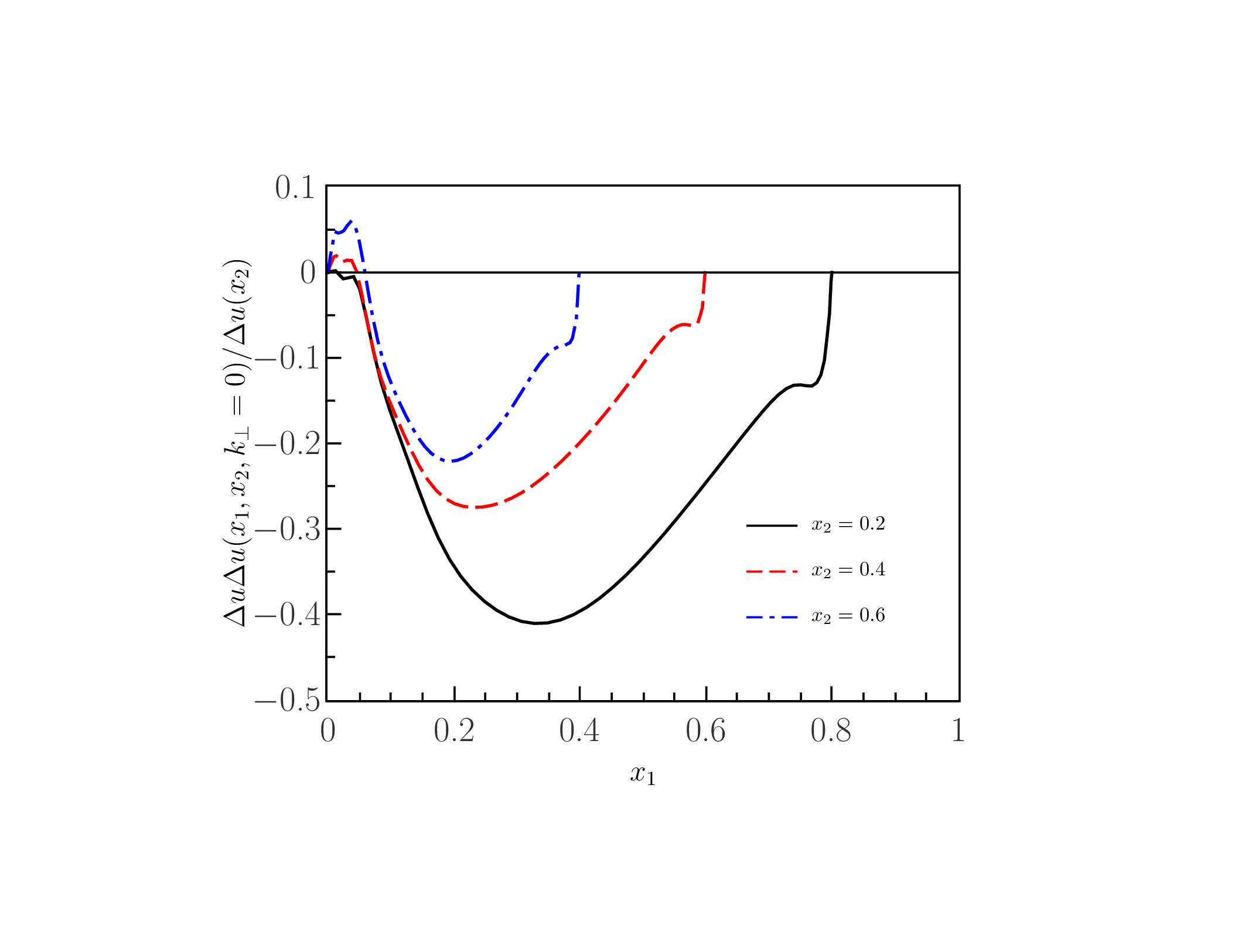}
\caption{\footnotesize The ratio $r_2$, Eq. (\ref{r2}), at $k_\perp =0$,
evaluated for three values 
of $x_2$, as a function of $x_1$.}
\end{minipage}
\end{figure}

\begin{figure}[t]
\begin{minipage}[t] {70 mm}
\vspace{7.0cm}
\includegraphics{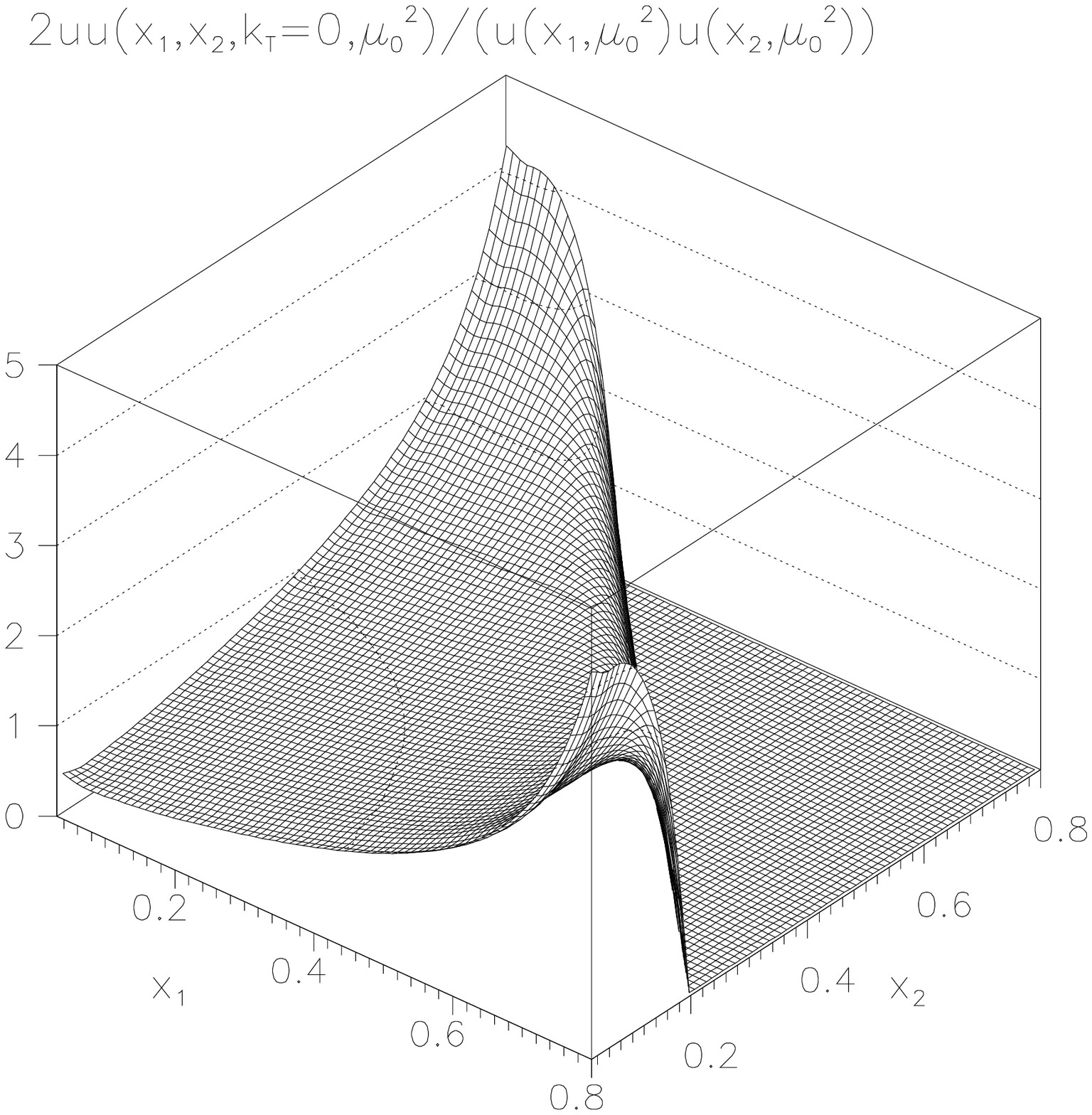}
\caption{\footnotesize  The ratio Eq. (\ref{r3}) at the hadronic scale.}
\end{minipage}
\hspace{\fill}
\begin{minipage}[t] {70 mm}
\vspace{7.0cm}
\includegraphics{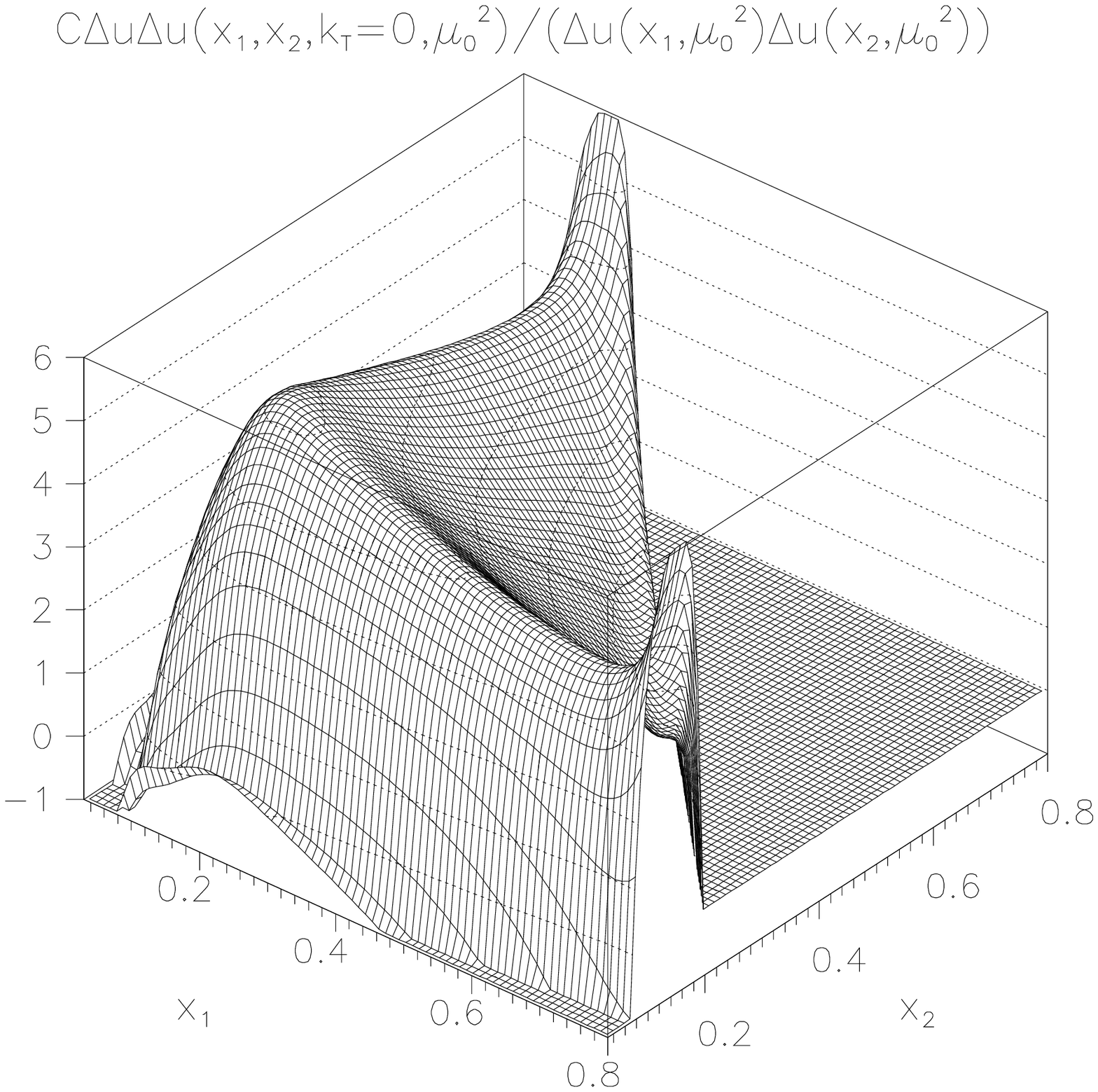}
\caption{\footnotesize The ratio Eq. (\ref{r4}) at the hadronic scale.}
\end{minipage}
\end{figure}

The numerical results for the calculation of the unpolarized and
longitudinally polarized dPDFs
at the hadronic scale of
the model, 
performed using this momentum space wave function in Eq. (\ref{uuspin}),
for example
in the case of two quarks of
flavor $q_1 = u, ~ q_2 = u$, 
are now described and discussed.
First of all, let us remark that, in proper limits,
the calculated unpolarized dPDFs
reproduce the single particle PDFs evaluated in the 
same LF hypercentral scheme and shown in Ref. \cite{pasquini}.

In Figs. 1 and 2 the
distributions
$uu(x_1,x_2, k_\perp =0)$ and $\Delta u \Delta u (x_1, x_2, k_\perp =0)$
are shown for three different values of $x_2$. 
At variance with
the results obtained in previous works, Refs. \cite{nostro,
Manohar:DPS2}, a correct 
support is obtained, i.e., the dPDFs are different from zero
only in the physical
region, $x_1+x_2<1$.
Thanks to the same property,
the symmetry of the dPDF in exchanging
$x_1$ and $x_2$, due to the
indistinguishability of the two quarks when $k_\perp =0$, 
is restored and the probabilistic 
interpretation of the distribution 
is recovered. 
In particular, for the unpolarized dPDF, where the SU(6)
spin-isospin symmetry breaking is not apparent as it is
in the polarized case due
to the presence of the Melosh rotation, one finds the following condition:

\begin{eqnarray}
uu(x_1,x_2, k_\perp=0)=uu(1-x_1-x_2,x_2, k_\perp=0)~.
\end{eqnarray}

This equation reflects the fact that,
once  the isospin factor coming
from the SU(6) symmetry has been taken into
account in the
calculation, the dPDF is completely symmetric under the exchange of any pair
of the three quarks. 

As already pointed out,
the main interest in this work is to understand the dynamical origin of the
correlations, and in particular to verify if the factorized
{\it ans\"atze} of  Eqs.
(\ref{app2},\ref{app1})
are valid in the valence quark region, where the used
CQM could give predictions on the distributions of interest here. 
To this aim, in order to estimate how much this factorization 
could be violated, it is instructive to analyze and show the ratios

\begin{eqnarray}
\label{r1}
 r_1(x_1,x_2) = \dfrac{uu(x_1,x_2, k_\perp=0)}{u(x_2)}~,
\end{eqnarray}

\begin{eqnarray}
\label{r2}
 r_2(x_1,x_2) = \dfrac{\Delta u \Delta u(x_1,x_2, k_\perp=0)}
{\Delta u(x_2)}~,
\end{eqnarray}

where $u(x_2)$ and $\Delta u(x_2)$ are the standard one-body unpolarized
and polarized PDFs, respectively.
The above ratios are shown in Figs. 3 and 4. If the
approximation
Eq. (\ref{app1}) were valid, the above ratios
should not depend on $x_2$. On the contrary, it is clearly seen
that in both the polarized and unpolarized case the
factorization is strongly violated, due to the correlations between the two
active quarks, present in the hypercentral CQM under scrutiny here.
These results are qualitatively in agreement with those found in Refs.
\cite{nostro,Manohar:DPS2}; of course, the amount of violation of 
the factorization and
the shape of the ratio are model dependent features. 

A useful pictorial representation of this 
analysis can be obtained by drawing two other
ratios,
\begin{eqnarray}
\label{r3}
 r_3(x_1,x_2) = \dfrac{2uu(x_1,x_2, k_\perp =0)}{u(x_1)u(x_2)}
\end{eqnarray}
and
\begin{eqnarray}
\label{r4}
 r_4(x_1,x_2) = \dfrac{ C \Delta u \Delta u(x_1,x_2, k_\perp =0)}
{\Delta u(x_1) \Delta u(x_2)}~,
\end{eqnarray}
where $C$ is the constant
\begin{eqnarray}
\label{cost}
C = \dfrac{ [ \int dx_1 \Delta u(x_1) ]^2}
{\int dx_1 \, dx_2\,\Delta u \Delta u(x_1, x_2, k_\perp=0)}~,
\end{eqnarray}
in a three-dimensional plot.
The factors 2 and $C$ in the numerator of 
Eqs. (\ref{r3}) and (\ref{r4}), respectively,
have been added
in order to have these ratios 
equal to 1 in the kinematical regions where the
factorization {\it ansatz} 
Eq. (\ref{app1}) is valid.
Clearly, the constant $C$ is a model dependent quantity and its value
is $8/3$ in a pure NR SU(6) scheme and $-6.17$ in the present LF 
hypercentral
approach.
The ratios Eqs. (\ref{r3}) and (\ref{r4}) are shown in Figs. 5 
and 6,
respectively. 
It is clear that in our approach the factorization property,
Eq. (\ref{app1}), 
is badly violated, in particular in the
polarized case where, due to the 
contribution of the Melosh
rotation, the correlations between the two quarks are so strong to lead
to a change of the sign of the dPDF with respect to the single particle
PDF, which determines a severe violation of the 
approximation, for any value of $x$.
It is interesting to notice that, if one had used a NR SU(6) scheme,
Fig. 5 and 6 would be exactly the same.
The shown difference 
is a model dependent, relativistic effect.
In the unpolarized case, the shape of the ratio Eq. (\ref{r3})
is indeed not too different from that obtained in \cite{nostro}
in a conventional CQM approach.
In the polarized case, relativity produces a big difference.
This seems to indicate that the factorization ansatz should be used
with great care in the spin-dependent case. 
\begin{figure}[t]
\begin{minipage}[t] {70 mm}
\vspace{7.0cm}
\includegraphics{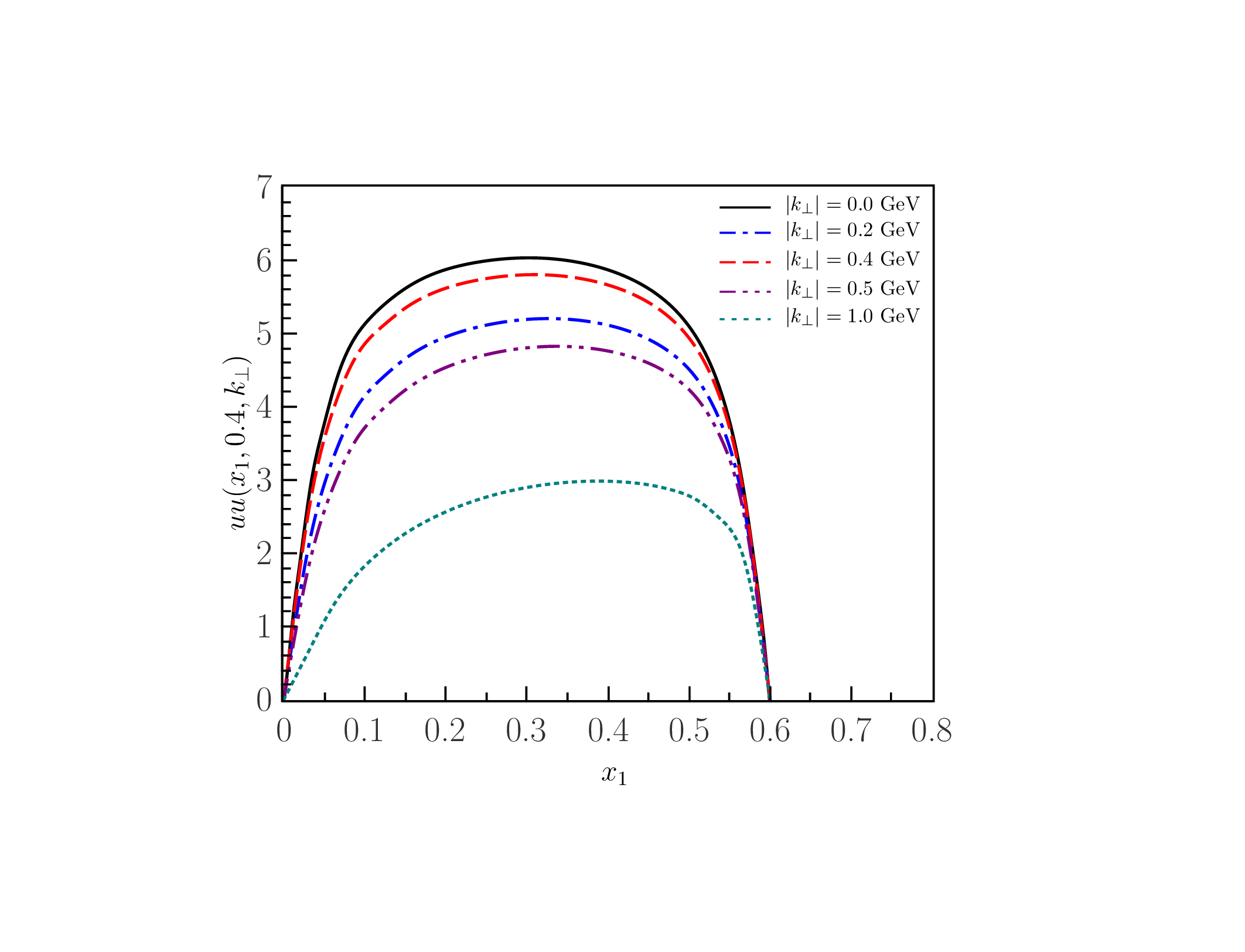}
\caption{\footnotesize The unpolarized dPDF, Eq. (\ref{unp}), evaluated at
$x_2 = 0.4$ and at five values of $k_\perp$. }
\end{minipage}
\hspace{\fill}
\begin{minipage}[t] {70 mm}
\vspace{7.0cm}
\includegraphics{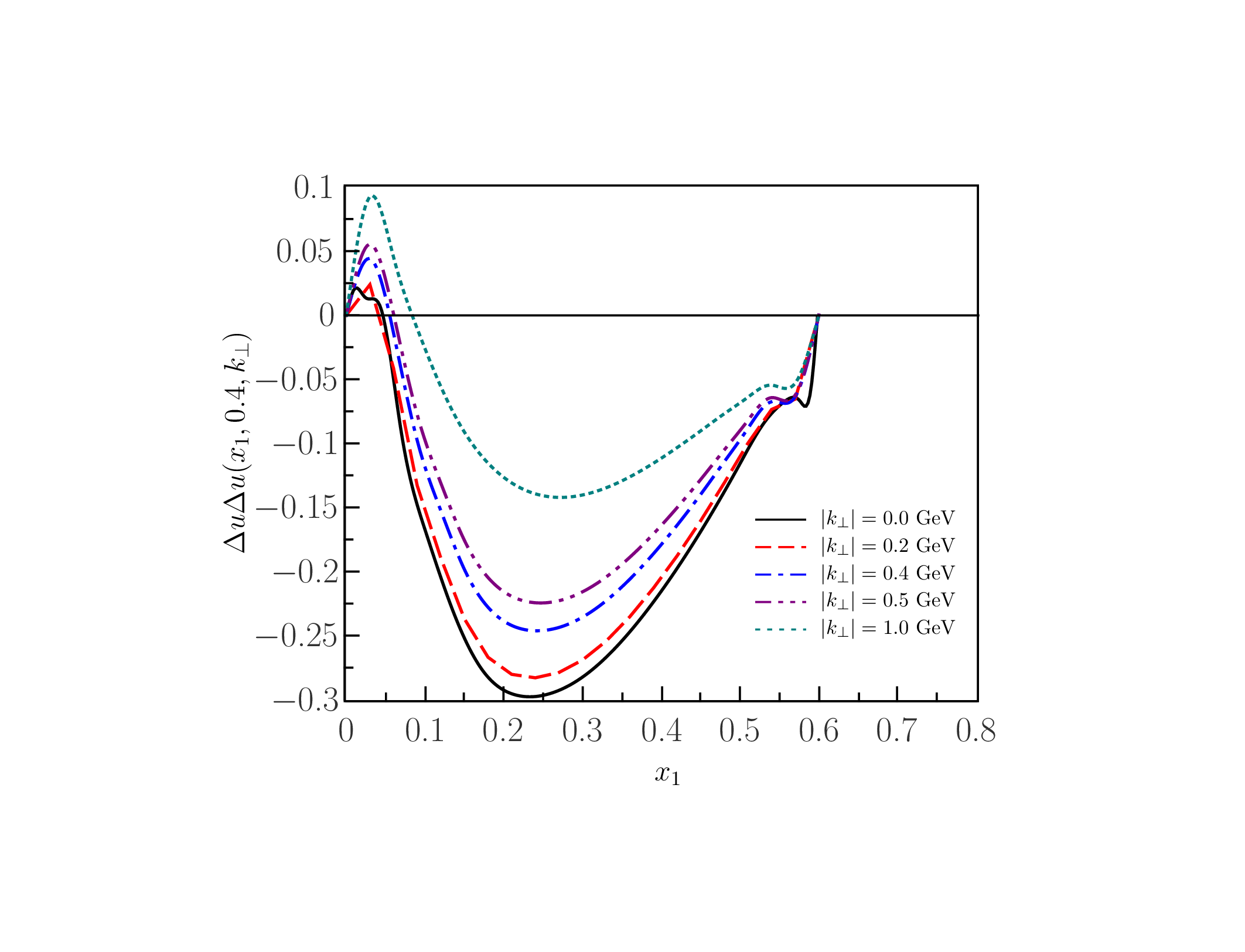}
\caption{ \footnotesize The 
polarized dPDF, Eq. (\ref{pol}), evaluated at
$x_2 = 0.4$ and at five values of $k_\perp$.}
\end{minipage}
\end{figure}

\begin{figure}[t]
\begin{minipage}[t] {70 mm}
\vspace{7.0cm}
\includegraphics{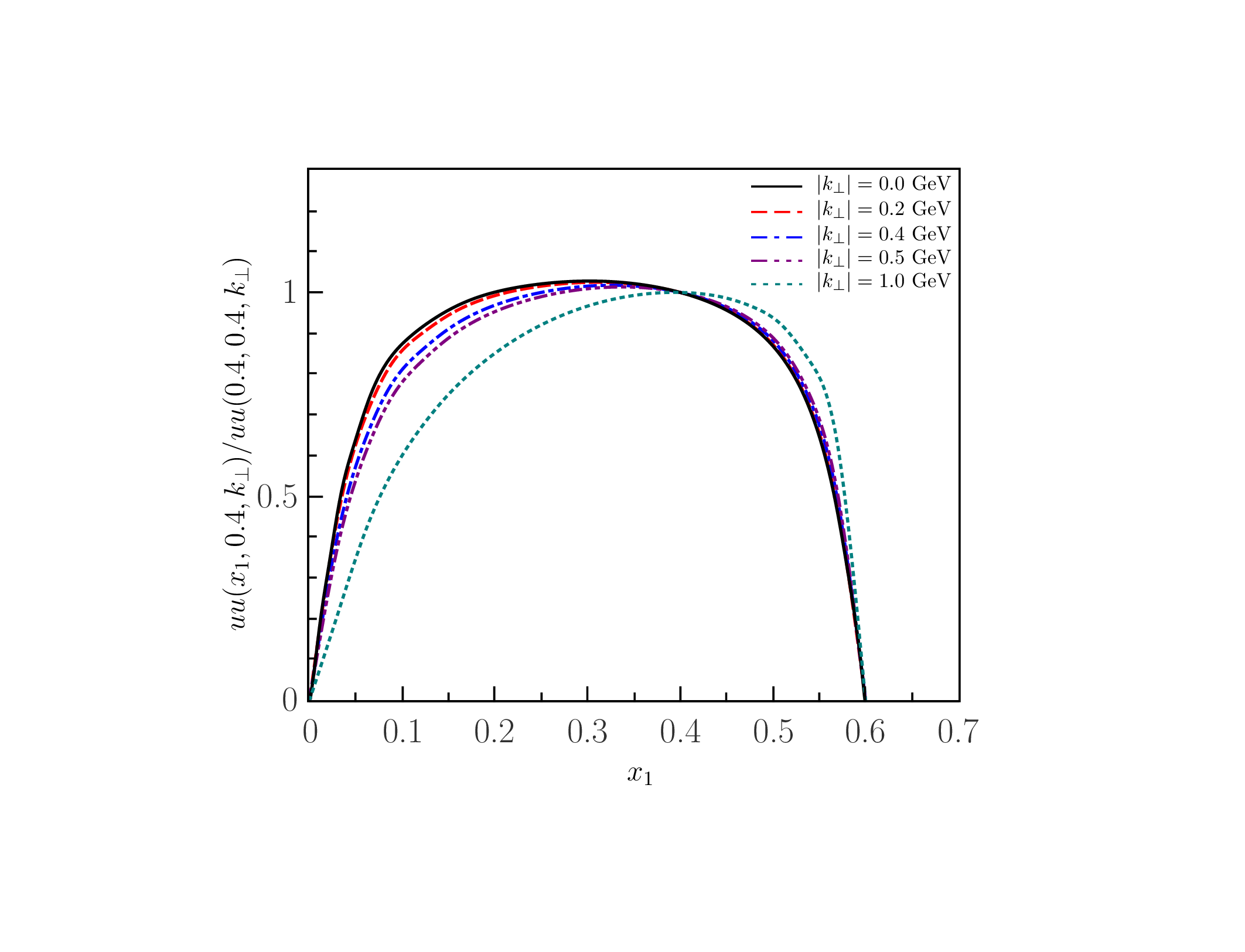}
\caption{ \footnotesize The ratio $r_5$, Eq. (\ref{r5}), for
five values of $k_\perp$.}
\end{minipage}
\hspace{\fill}
\begin{minipage}[t] {70 mm}
\vspace{7.0cm}
\includegraphics{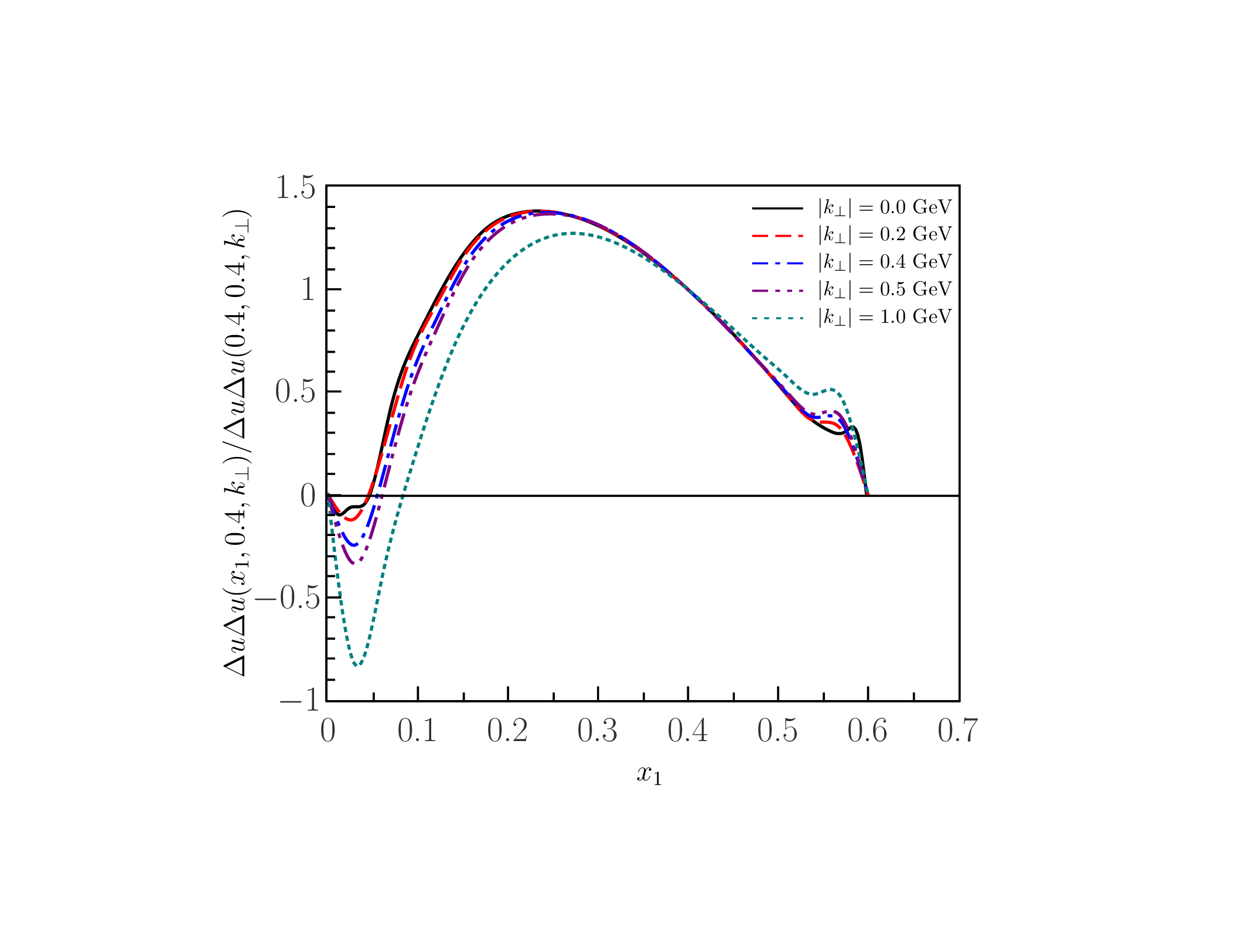}
\caption{\footnotesize The ratio $r_6$, Eq. (\ref{r6}), 
for five values of $k_\perp$.}
\end{minipage}
\end{figure}

In Figs. 7 and 8, the polarized and unpolarized
dPDFs 
are respectively shown, for five values
of the difference in transverse momentum, $k_\perp$. 
The behavior of the dPDFs, decreasing with increasing  
$k_\perp$, is similar to that described in
Refs. \cite{nostro, Manohar:DPS2}.
A better insight in the $k_\perp$ dependence is seen
in Figs. 9 and 10, where the ratios

\begin{eqnarray}
\label{r5}
r_5(x_1,x_2,k_\perp)= \dfrac{uu(x_1,0.4, k_\perp)}{uu(0.4,0.4, k_\perp)}
\end{eqnarray}

and

\begin{eqnarray}
\label{r6}
r_6(x_1,x_2,k_\perp)= \dfrac{\Delta u
\Delta u (x_1,0.4, k_\perp)}{\Delta u \Delta u(0.4,0.4, k_\perp)}~,
\end{eqnarray}
 
are shown for five values
of $k_\perp$. This quantity, already analyzed in Refs.
\cite{nostro, Manohar:DPS2},
has been chosen in order to check the validity of
the approximation Eq. (\ref{app1}), i.e.,
the possibility to factorize  
the dependences on $x_1,x_2$ and $k_\perp$ 
in the dPDFs. 
The violation of this {\it ansatz} grows smoothly 
with increasing $k_\perp$. 
{ In the present LF approach, this is somehow expected.
First of all, since the invariant mass Eq. (\ref{invariant}) 
is defined in terms of both $x$ and $k_\perp$,
the dependence of a LFWF on these variables cannot be separated.
This is relevant, for example, in studies of
leading-twist lensing effects (see, e.g., Ref. \cite{kperp}).  
The breaking of the $x$ and $k_\perp$ factorization  
is correctly found in our model LFWFs. 
Separable forms (often invoked)
are rejected by the structure of our relativistically
invariant approach.
Moreover, in our scheme,
the relativistic effect provided by the Melosh rotations, 
which are present also in the unpolarized case, when $k_\perp \ne 0$,
is another source of unfactorized $x$ and $k_\perp$ dependencies.
It could be worth to remind instead that,
in a pure SU(6) NR scenario, the factorization holds \cite{nostro}.} 

In closing this section 
we notice that, to have a flavor
of the behavior of the dPDFs at high momentum transfer and, possibly, in
the low-$x$ kinematical region,
$x \lesssim 10^{-2}$, 
presently accessible at the LHC, the pQCD evolution
of the model results is necessary.
An analysis of this kind, i.e. the check
of the validity of the approximations Eqs.
(\ref{app2},\ref{app1}) at high momentum scales,
is reported in the next section.


\section{pQCD evolution of the model results}

In this section, the procedure and the results 
of the pQCD evolution of the 
model calculation will be described. For the moment being, 
evolved unpolarized and polarized dPDFs are presented
at Leading-Order (LO),
only when the transverse distance between
the two partons in momentum space is zero, i.e., $k_\perp = 0$.
Besides, as everywhere in the paper,
the scale has been taken to be the same for both partons, which
is not the case in actual processes, in general.

The evolution equations for the dPDFs, introduced in Refs. 
~\cite{Kirschner,Shelest:1982dg}
and later used and discussed in Refs.
~\cite{Diehl:2011yj,Manohar:2012jr,
Cattaruzza:2005nu, 
Diehl:2011tt,agg1,agg2,Ceccopieri:2010kg,
agg3,agg4,Manohar:DPS2,arriola,kasemets2,
snigirev_evo},
are a generalization of the DGLAP 
equations so that, also in this case, the solution can be found by using 
the Mellin transformations of the calculated functions.
Since we analyse only the valence quark contribution, the 
inhomogeneous part of the evolution equations is not involved (see also 
Refs. \cite{arriola,kasemets2,snigirev_evo}). Thanks to this fact, 
the evolved dPDFs can be found as follows.

We use the Mellin moments of the weighted distributions:
\begin{eqnarray}
&& x_1 x_2 F_{i_1,i_2}(x_1,x_2,Q^2) \,, \nonumber\\
&& x F_i(x,Q^2) \,,
\end{eqnarray}
i.e., explicitly 
\begin{eqnarray}
\langle xF_i(Q^2) \rangle_n & = & \int_0^1 dx\, x^{n-2}\,x F_i(x,Q^2)
\,,\nonumber\\
\langle x_1 x_2 F_{i_1,i_2}(Q^2) 
\rangle_{n_1,n_2} &=& \int_0^1 d x_1 \int_0^1 d x_2 \,x_1^{n_1 -2}\, 
x_2^{n_2 - 2}\,x_1 x_2 F_{i_1,i_2}(x_1,x_2,Q^2)\,.
\end{eqnarray}
Here a somehow shorthand notation is adopted,
where the PDF for a quark of flavour $i$  
is named $F_i(x)$ and the
dPDFs for two polarized or unpolarized 
quarks of flavours $i_1$ and $i_2$, 
is 
$F_{i_1 i_2}(x_1,x_2, k_\perp=0) = F_{i_1 i_2}(x_1,x_2)$.
Their LO evolution reads
\begin{eqnarray}
\langle xF_i(Q^2) \rangle _ n & = & 
\left({a_s \over a_{s0}}\right)^{-{P^{(0)}_{\rm NS}(n)\over \beta_0}} \, 
\langle x F_i(\mu_0^2) \rangle \,,\nonumber\\
 \langle x_1 x_2 F_{i_1,i_2}(Q^2) \rangle_{n_1,n_2} &=& 
\left({a_s \over a_{s0}}\right)^{-{P^{(0)}_{\rm NS}(n_1)\over \beta_0}} 
\cdot  \left({a_s \over a_{s0}}\right)^{-{P^{(0)}_{\rm NS}(n_2)\over \beta_0}} 
\,\, \langle x_1 x_2 F_{i_1,i_2}(\mu_0^2) \rangle_{n_1,n_2} \,
\label{eq:splitting-fs}
\end{eqnarray}
where: 
$a_{s0} = {\alpha_s(\mu^2_0) \over 4 \pi}$ and 
$a_{s} = {\alpha_s(Q^2) \over 4 \pi}$
and, at LO,
$
 a_s \equiv a_{s,{\rm LO}} =  {1 / (\beta_0 \ln(Q^2/\Lambda_{\rm 
LO}^2))}\,,
$
$\beta_0 = 11 - 2 n_f/3$. 
$P^{(0)}_{\rm NS}(n)$ 
is $n^{th}$ moment of the Non-Singlet splitting function at LO.
Taking the $n$, $n_1$ and $n_2$ moments complex, the evolved distributions
in $x$-space
are obtained in terms of the
inverse Mellin transformations.
Namely
\begin{eqnarray}
x F_{i}(x,Q^2) & = &{1 \over 2 \pi i}  \oint_{\cal C} dn \, x^{(1-n)}\, \langle xF_{i}(Q^2)\rangle_n = \nonumber \\
x_1 x_2  F_{i_1,i_2}(x_1,x_2,Q^2) & = & 
{1 \over 2 \pi i} \oint_{\cal C} dn_1 \,
{1 \over 2 \pi i} \oint_{\cal C} dn_2 \,x_1^{(1-n_1)}\, x_2^{(1-n_2)}\, 
\langle x_1 x_2  F_{i_1,i_2}(Q^2) \rangle_{n_1,n_2}\,.
\end{eqnarray}

We have implemented the numerical evaluation of the dPDFs evolution.
Our code reproduces correctly the results shown in Ref.
\cite{arriola}, if we use as an input the simple factorized {\it ansatz}
used in that paper. The good support property of the present LF calculation
helps to obtain a proper evaluation of the Mellin moments
for a precise evolution procedure.
\begin{figure}[t]
\vspace{15.0cm}
\includegraphics{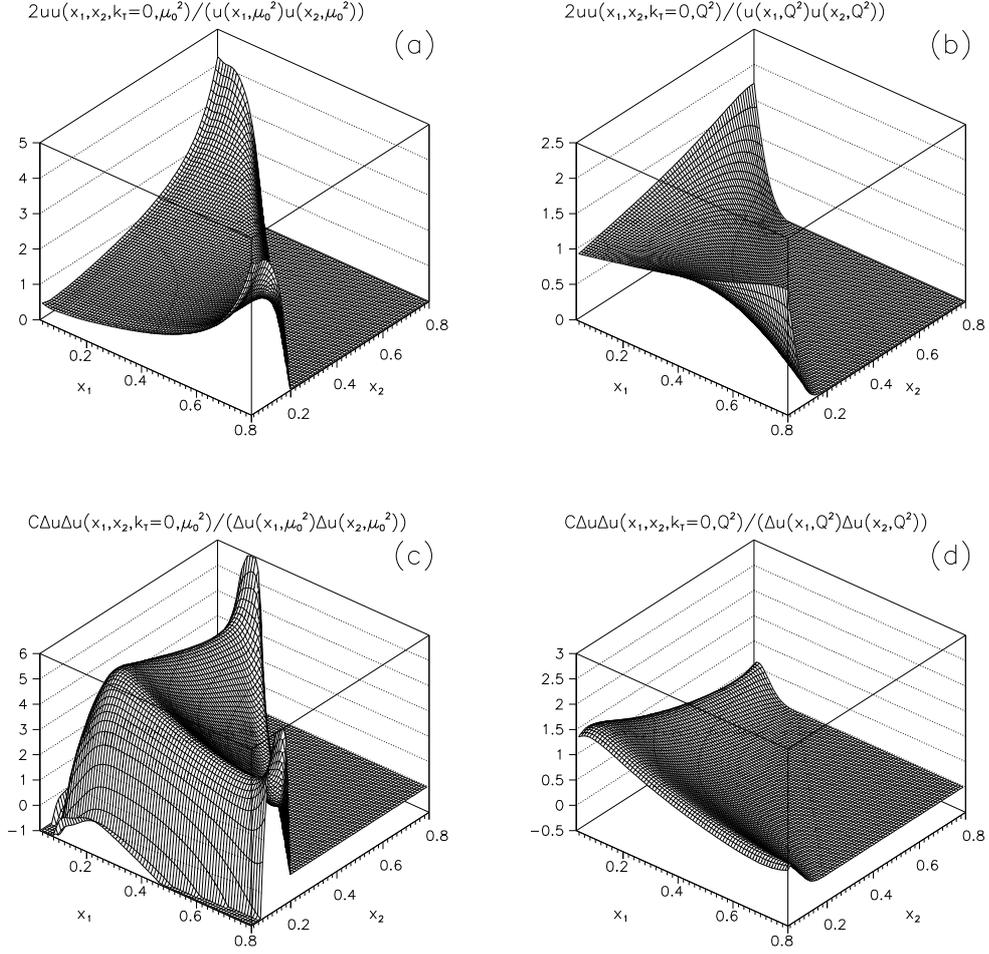}
\caption{ \footnotesize a) The ratio Eq. (\ref{r3}) at the hadronic scale;
b) the same quantity at a scale $Q^2= 10$ GeV$^2$;
c) the ratio Eq. (\ref{r4}) at the hadronic scale;
d) this last quantity at a scale $Q^2= 10$ GeV$^2$.
The vertical scale of panels (b) and (d) is reduced by a factor of 2
with respect to panels (a) and (c), respectively.} 
\end{figure}

The present calculation scheme does not fix the hadronic scale $\mu_o^2$.
To this aim, we have followed here the proposal 
of, e.g., Ref. \cite{trvlc}: $\mu_o^2$
is the scale at which
the valence quarks of the model carry all the proton momentum,
i.e., the second moment of the unpolarized valence parton distribution is 1.
Knowing the experimental value of this quantity
at a high momentum scale,
one can evolve it using LO pQCD to lower scales.
$\mu_o^2$ is the scale at which the second moment of the unpolarized
valence parton distribution turns out to be one.
In this way, a very low value for $\mu_o^2$ is found.
In agreement with Ref. \cite{trvlc},
we fixed the hadronic scale to the value of 0.08 GeV$^2$,
consistent with that used in the analyses shown, e.g., in Refs. 
\cite{vlc_pg,h1,siv_bm,LF2,pasquevo,marco1,marco2} for PDFs and GPDs.
A very low starting scale has been tested also for dPDFs
in Ref. \cite{kasemets2}.
Actually, recent studies have proposed, 
using different arguments, that models
can be associated to a little higher scales, which would make a difference
in the numerical results
(see, e.g., \cite{pasqmu}). Nevertheless we are performing here an exploratory
calculation since there are no available data for these observables.
We think that
we should not care of this aspect for the moment being.
For the same reason we have taken a standard final scale, $Q^2$ = 10
GeV$^2$, although the one at the LHC is certainly larger.
We think in fact that an evolution from
0.08 to 10 GeV$^2$ is extreme enough to simulate qualitatively
the change in reaching, through pQCD, the experimental situation
from a model calculation. 

In Figs. 11 (b) and 11 (d) the results of this procedure are shown
for the ratios Eqs. (\ref{r3}) and (\ref{r4}),
starting from the model distributions at the hadronic scale giving
the ratios in Figs. 11 (a) and 11 (c),
respectively, already discussed in the previous section and
presented here again for readers convenience.  
Now, Fig. 11 provides a complete summary of the results of 
the present analysis.
All the shown ratios are constructed to yield one if correlations are not
active.
It is clearly seen that, in the unpolarized case, 
the effect of correlations 
keeps being sizable after evolution
in the valence $x$ region. At the same time,
it gets less important at small $x$, where it is
investigated presently at the LHC, as already noticed in
Refs. \cite{kasemets2,arriola,snigirev_evo} using simple
test functions as inputs in the evolution equations.
However, in the model under scrutiny, it is found that
spin-dependent correlations, present also
in the case of scattering of
unpolarized protons, are important, after evolution, even at low $x$.

In closing, we note that
a realistic evaluation of DPS at low $x$ would require
a complete study of the evolution of dPDFs, considering therefore
not only the valence, non-singlet sector but
also the produced gluons and sea quarks, which are
mixing in the evolution of the singlet sector.
For the moment being, the results reported here are intended to give
a first useful glance on the general effects of the evolution.

\section{Conclusions}

The 
double parton distribution functions (dPDFs) have been evaluated,
in the valence sector,
in both the spin-independent and spin dependent case,
accessible, in principle, in the scattering of unpolarized protons occurring
at, e.g., the LHC.
The chosen framework has been a Light-Front constituent quark model.
The main aim was to check to what extent constituent quark model
estimates support the factorization {\it ansatz} often used 
in phenomenological applications.
According to this procedure
parton correlations are neglected, e.g.,
dPDFs are factorized in two terms, 
one depending on the longitudinal
momentum fractions of the two quarks, $x_1$ and $x_2$, and another 
on the transverse momentum separation, $\vec k_\perp$, e.g. a sort
of $x - \vec k_\perp$ factorization is assumed.
In addition to that, also an $x_1-x_2$ factorization assumption
is made, e.g., the distribution
on the longitudinal momenta $x_1, x_2$ is taken to be uncorrelated.

Previous quark model analyses have been performed in
a properly modified version of the standard uncorrelated
bag model and in a Non-Relativistic constituent quark model,
where correlations are present from the very beginning.
The outcome was that, in both models,
the $x_1-x_2$ factorization is strongly violated,
while the $x - k_\perp$ factorization is mildly violated.
These results have been tested here within a relativistic, 
fully Poincar\'e covariant
Light-Front approach, already used 
for the calculation of single particle parton distributions. 
As a consequence of the Poincar\'e covariance,
some symmetries expected on general grounds, lost in previous
analyses, are clearly recovered and
the so called
``bad support'' problem, i.e., the fact that 
dPDFs are not vanishing in the forbidden
kinematic region, $x_1+x_2 > 1$, does not arise.
The main results are as follows.
At the low momentum scale of the model, the strong violation of
the $x_1-x_2$ factorization and the mild one of
the $x - k_\perp$ factorization are basically confirmed.
The obtained valence dPDFs have been then pQCD evolved
to a high momentum scale, to give an idea on what could happen
in the experimentally accessible region.
After pQCD evolution the situation is somehow more subtle. 
While the effect of correlations 
keeps being sizable after evolution
in the valence $x$ region,
for the unpolarized distribution
it gets less important at small $x$.
However, spin-dependent correlations, present also
in the case of scattering of
unpolarized protons, are not washed out
by the evolution and are sizable even at low $x$ values.

Further studies, including the evolution of the singlet sector, important
for the description of the low $x$ region presently observed at the LHC,
and the evaluation of the contributions of DPS to the cross sections of
specific relevant channels, are in progress. 
{ Besides, from the formal point of view, our LF approach to double 
parton correlations could be tested by studying the
cluster decomposition properties at weak binding of 
the LFWFs.  This is a relevant feature that correctly defined LFWFs 
should show  \cite{cluster}. 
Another way to access parton correlations
in the free proton is the analysis of interference effects in proton
nucleus collisions, firstly addressed in \cite{trelest}.
Calculations of interference effects generalizing the approach recently used
in Ref. \cite{simona} are going on.}

\section{Acknowledgments}

This work was supported in part by the Research Infrastructure
Integrating Activity Study of Strongly Interacting Matter (acronym
HadronPhysic3, Grant Agreement n. 283286) 
under the Seventh Framework
Programme of the European Community,
by the Mineco
under contract FPA2010-21750-C02-01,
by GVA-Prometeo/2009/129, and by CPAN(CSD-00042).
S.S. thanks the Department
of Theoretical Physics of the University of Valencia for warm hospitality
and support.
M.T. and V.V. thank the INFN, sezione di Perugia and the Department
of Physics of the University of Perugia for warm hospitality
and support.


\begin{thebibliography}{34}%

\bibitem{Paver:1982yp}
  N.~Paver and D.~Treleani,
``Multi - Quark Scattering And Large P(t) Jet Production In Hadronic 
Collisions",
Nuovo Cim.\ A {\bf 70} (1982) 215.

\bibitem{livio} 
  T.~Akesson {\it et al.}  [Axial Field Spectrometer Collaboration],
 ``DOUBLE PARTON SCATTERING IN p-p COLLISIONS AT S**(1/2) = 63-GeV",
  Z.\ Phys.\ C {\bf 34}, 163 (1987).

\bibitem{Gaunt:2009re} 
  J.~R.~Gaunt and W.~J.~Stirling,
``Double Parton Distributions Incorporating Perturbative QCD Evolution and 
Momentum and Quark Number Sum Rules",
  JHEP {\bf 1003}, 005 (2010)
[arXiv:0910.4347 [hep-ph]]~.

\bibitem{Diehl:2011yj} 
  M.~Diehl, D.~Ostermeier and A.~Schafer,
``Elements of a theory for multiparton interactions in QCD",
JHEP {\bf 1203}, 089 (2012)
[arXiv:1111.0910 [hep-ph]]~.

\bibitem{Manohar:2012jr} 
  A.~V.~Manohar and W.~J.~Waalewijn,
``A QCD Analysis of Double Parton Scattering: Color Correlations, 
Interference Effects and Evolution",
  Phys.\ Rev.\ D {\bf 85}, 114009 (2012)
[arXiv:1202.3794 [hep-ph]]~.

\bibitem{pg}
  P.~Bartalini, (ed.) and L.~Fan\`o, (ed.),
``Multiple partonic interactions at the LHC. Proceedings, 1st International 
Workshop, MPI'08, Perugia, Italy, October 27-31, 2008",
arXiv:1003.4220 [hep-ex].

\bibitem{Kulesza:1999zh} 
  A.~Kulesza and W.~J.~Stirling,
``Like sign $W$ boson production at the LHC as a probe of double parton 
scattering",
  Phys.\ Lett.\ B {\bf 475}, 168 (2000) [hep-ph/9912232].

\bibitem{Cattaruzza:2005nu} 
  E.~Cattaruzza, A.~Del Fabbro and D.~Treleani,
``Fractional momentum correlations in multiple production of $W$ bosons and 
of $b \bar{b}$ pairs in high energy $p p$ collisions",
  Phys.\ Rev.\ D {\bf 72}, 034022 (2005)  [hep-ph/0507052].

\bibitem{Maina:2009sj} 
  E.~Maina,
  JHEP {\bf 0909}, 081 (2009) [arXiv:0909.1586 [hep-ph]].

\bibitem{Gaunt:2010pi} 
  J.~R.~Gaunt, C.~-H.~Kom, A.~Kulesza and W.~J.~Stirling,
``Same-sign W pair production as a probe of double parton scattering at the 
LHC",
  Eur.\ Phys.\ J.\ C {\bf 69}, 53 (2010) [arXiv:1003.3953 [hep-ph]].

\bibitem{Kasemets:2012pr} 
  T.~Kasemets and M.~Diehl,
``Angular correlations in the double Drell-Yan process",
  JHEP {\bf 1301}, 121 (2013) [arXiv:1210.5434 [hep-ph]]~.

\bibitem{DelFabbro:1999tf} 
  A.~Del Fabbro and D.~Treleani,
``A Double parton scattering background to Higgs boson production at the 
LHC",
Phys.\ Rev.\ D {\bf 61}, 077502 (2000) [hep-ph/9911358].

\bibitem{Hussein:2006xr}
  M.~Y.~Hussein,
``A Double parton scattering background to associate WH and ZH production 
at the LHC",
  Nucl.\ Phys.\ Proc.\ Suppl.\  {\bf 174} (2007) 55 [hep-ph/0610207]~.

\bibitem{Bandurin:2010gn} 
  D.~Bandurin, G.~Golovanov and N.~Skachkov,
``Double parton interactions as a background to associated HW production at 
the Tevatron", JHEP {\bf 1104}, 054 (2011) [arXiv:1011.2186 [hep-ph]].

\bibitem{Berger:2011ep} 
  E.~L.~Berger, C.~B.~Jackson, S.~Quackenbush and G.~Shaughnessy,
  Phys.\ Rev.\ D {\bf 84}, 074021 (2011) [arXiv:1107.3150 [hep-ph]].

\bibitem{Aad:2013bjm} 
  G.~Aad {\it et al.}  [ATLAS Collaboration],
New J.\ Phys.\  {\bf 15}, 033038 (2013) [arXiv:1301.6872 [hep-ex]].


\bibitem{Mekhfi:1985dv} 
  M.~Mekhfi,
``Correlations In Color And Spin In Multiparton Processes",
  Phys.\ Rev.\ D {\bf 32}, 2380 (1985).

\bibitem{Diehl:2011tt} 
  M.~Diehl and A.~Schafer,
``Theoretical considerations on multiparton interactions in QCD",
  Phys.\ Lett.\ B {\bf 698}, 389 (2011) [arXiv:1102.3081 [hep-ph]].



\bibitem{weiss} 
  P.~Schweitzer, M.~Strikman and C.~Weiss,
  ``Intrinsic transverse momentum and parton correlations from dynamical 
chiral symmetry breaking",
  JHEP {\bf 1301}, 163 (2013) [arXiv:1210.1267 [hep-ph]].

\bibitem{oggi}
  M.~Diehl and T.~Kasemets,
  ``Positivity bounds on double parton distributions",
  JHEP {\bf 1305} (2013) 150 [arXiv:1303.0842 [hep-ph]].

\bibitem{pape}
  G.~Parisi and R.~Petronzio,
  ``On the Breaking of Bjorken Scaling",
  Phys.\ Lett.\ B {\bf 62} (1976) 331.

\bibitem{jaro}
  R.~L.~Jaffe and G.~G.~Ross,
  ``Normalizing the Renormalization Group Analysis of Deep Inelastic 
Leptoproduction",
  Phys.\ Lett.\ B {\bf 93} (1980) 313.

\bibitem{Kirschner} 
  R.~Kirschner,
  ``Generalized Lipatov-altarelli-parisi Equations And Jet Calculus Rules",
  Phys.\ Lett.\ B {\bf 84}, 266 (1979).

\bibitem{Shelest:1982dg} 
  V.~P.~Shelest, A.~M.~Snigirev and G.~M.~Zinovev,
  ``The Multiparton Distribution Equations In Qcd",
  Phys.\ Lett.\ B {\bf 113}, 325 (1982).


\bibitem{agg1} 
  M.~G.~Ryskin and A.~M.~Snigirev,
  ``A Fresh look at double parton scattering",
  Phys.\ Rev.\ D {\bf 83}, 114047 (2011) [arXiv:1103.3495 [hep-ph]]~.


\bibitem{agg2} 
  J.~R.~Gaunt,
  ``Double parton scattering singularity in one-loop integrals",
  PoS RADCOR {\bf 2011}, 040 (2011) arXiv:1103.1888 [hep-ph]~.

\bibitem{Ceccopieri:2010kg} 
  F.~A.~Ceccopieri,
  ``An update on the evolution of double parton distributions,''
  Phys.\ Lett.\ B {\bf 697}, 482 (2011)
  [arXiv:1011.6586 [hep-ph]]~.

\bibitem{agg3} 
  J.~R.~Gaunt,
  ``Single Perturbative Splitting Diagrams in Double Parton Scattering",
  JHEP {\bf 1301}, 042 (2013) [arXiv:1207.0480 [hep-ph]]~.

\bibitem{agg4} 
  B.~Blok, Y.~Dokshitser, L.~Frankfurt and M.~Strikman,
  ``pQCD physics of multiparton interactions",
  Eur.\ Phys.\ J.\ C {\bf 72}, 1963 (2012) [arXiv:1106.5533 [hep-ph]]~.
 


\bibitem{Manohar:DPS2} 
  A.~V.~Manohar and W.~J.~Waalewijn,
  ``What is Double Parton Scattering?",
  Phys.\ Lett.\ B {\bf 713}, 196 (2012) [arXiv:1202.5034 [hep-ph]]~.

\bibitem{arriola}
  W.~Broniowski and E.~Ruiz Arriola,
  ``Valence double parton distributions of the nucleon in a simple model",
  Few Body Syst.\  {\bf 55}, 381 (2014) [arXiv:1310.8419 [hep-ph]]~.

\bibitem{kasemets2}
  M.~Diehl, T.~Kasemets and S.~Keane,
  ``Correlations in double parton distributions: effects of evolution",
  JHEP {\bf 1405}, 118 (2014) [arXiv:1401.1233 [hep-ph]]~.

\bibitem{snigirev_evo}
  A.~M.~Snigirev, N.~A.~Snigireva and G.~M.~Zinovjev,
  ``Perturbative and nonperturbative correlations in double parton 
distributions",
  Phys.\ Rev.\ D {\bf 90}, 014015 (2014) [arXiv:1403.6947 [hep-ph]]~.

\bibitem{man_bag}
  H.~-M.~Chang, A.~V.~Manohar and W.~J.~Waalewijn,
  ``Double Parton Correlations in the Bag Model",
  Phys.\  Rev.\  D 87, {\bf 034009} (2013) [arXiv:1211.3132 [hep-ph]]~.

\bibitem{Jaffe:1974nj} 
  R.~L.~Jaffe,
  ``Deep Inelastic Structure Functions in an Approximation to the Bag 
Theory",
  Phys.\ Rev.\ D {\bf 11}, 1953 (1975).

\bibitem{Benesh:1987ie} 
  C.~J.~Benesh and G.~A.~Miller,
  ``Deep Inelastic Structure Functions In The Mit Bag Model",
  Phys.\ Rev.\ D {\bf 36}, 1344 (1987).

\bibitem{nostro}
  M.~Rinaldi, S.~Scopetta and V.~Vento,
  ``Double parton correlations in constituent quark models",
  Phys.\ Rev.\ D {\bf 87} (2013) 11,  114021 [arXiv:1302.6462 [hep-ph]].

\bibitem{trvlc}
  M.~Traini, A.~Mair, A.~Zambarda and V.~Vento,
  ``Constituent quarks and parton distributions",
  Nucl.\ Phys.\  A {\bf 614}, 472 (1997).

\bibitem{h1}
  S.~Scopetta and V.~Vento,
  ``A quark model analysis of the transversity distribution",
  Phys.\ Lett.\  B {\bf 424}, 25 (1998) [arXiv:hep-ph/9706413]~.

\bibitem{orb}
  S.~Scopetta and V.~Vento,
  ``A quark model analysis of orbital angular momentum",
  Phys.\ Lett.\  B {\bf 460}, 8 (1999)
  [Erratum-ibid.\  B {\bf 474}, 235 (2000)] [arXiv:hep-ph/9901324]~.

\bibitem{vlc_pg}
  S.~Scopetta and V.~Vento,
  ``Generalized parton distributions in constituent quark models",
  Eur.\ Phys.\ J.\  A {\bf 16}, 527 (2003) [arXiv:hep-ph/0201265]~.

\bibitem{siv_bm}
  A.~Courtoy, F.~Fratini, S.~Scopetta and V.~Vento,
  ``A quark model analysis of the Sivers function",
  Phys.\ Rev.\  D {\bf 78} (2008) 034002 [arXiv:0801.4347 [hep-ph]]~.

\bibitem{siv_sr}
  A.~Courtoy, S.~Scopetta and V.~Vento,
 ``Model calculations of the Sivers function satisfying the Burkardt 
Sum Rule",
  Phys.\ Rev.\  D {\bf 79}, 074001 (2009) [arXiv:0811.1191 [hep-ph]]~.

\bibitem{bm_sr}
A.~Courtoy, S.~Scopetta and V.~Vento,
``Analyzing the Boer-Mulders function within different quark models",
  Phys.\ Rev.\ D {\bf 80}, 074032 (2009) [arXiv:0909.1404 [hep-ph]]~.

\bibitem{snig1} 
  V.~L.~Korotkikh and A.~M.~Snigirev,
  ``Double parton correlations versus factorized distributions",
  Phys.\ Lett.\ B {\bf 594}, 171 (2004) [hep-ph/0404155]~.

\bibitem{snig2} 
  A.~M.~Snigirev,
  ``Asymptotic behavior of double parton distribution functions",
  Phys.\ Rev.\ D {\bf 83}, 034028 (2011) [arXiv:1010.4874 [hep-ph]]~.

\bibitem{LF}
  B.~D.~Keister and W.~N.~Polyzou,
  ``Relativistic Hamiltonian dynamics in nuclear and particle physics",
  Adv.\ Nucl.\ Phys.\  {\bf 20} (1991) 225~.

\bibitem{LF0} 
  S.~J.~Brodsky, H.~-C.~Pauli and S.~S.~Pinsky,
  ``Quantum chromodynamics and other field theories on the light cone",
  Phys.\ Rept.\  {\bf 301}, 299 (1998) [hep-ph/9705477]~.

\bibitem{LF1} 
  F.~Cardarelli, E.~Pace, G.~Salm\`e and S.~Simula,
  ``Nucleon and pion electromagnetic form-factors in a light front 
constituent quark model",
  Phys.\ Lett.\ B {\bf 357}, 267 (1995) [nucl-th/9507037]~.

\bibitem{LF2} 
  P.~Faccioli, M.~Traini and V.~Vento,
  ``Polarized parton distributions and light front dynamics",
  Nucl.\ Phys.\ A {\bf 656} (1999) 400 [hep-ph/9808201]~.



\bibitem{pasquini}
  S.~Boffi, B.~Pasquini and M.~Traini,
  ``Linking generalized parton distributions to constituent quark models",
  Nucl.\ Phys.\ B {\bf 649}, 243 (2003) [hep-ph/0207340]~.



  
  
\bibitem{pasquinipol}
  S.~Boffi, B.~Pasquini and M.~Traini,
  ``Helicity dependent generalized parton distributions in constituent 
quark models",
  Nucl.\ Phys.\ B {\bf 680} (2004) 147 [hep-ph/0311016]~.

\bibitem{pasquevo}
  B.~Pasquini, M.~Traini and S.~Boffi,
  ``Nonperturbative versus perturbative effects in generalized parton 
distributions",
  Phys.\ Rev.\ D {\bf 71} (2005) 034022 [hep-ph/0407228]~.

\bibitem{pasquinih}
  B.~Pasquini, M.~Pincetti and S.~Boffi,
  ``Chiral-odd generalized parton distributions in constituent quark 
models",
  Phys.\ Rev.\ D {\bf 72}, 094029 (2005) [hep-ph/0510376]~.

\bibitem{marco1}
  M.~Traini,
  ``Charge symmetry violation: A NNLO study of partonic observables",
  Phys.\ Lett.\ B {\bf 707} (2012) 523 [arXiv:1110.3594 [hep-ph]]~.

\bibitem{marco2}
  M.~Traini,
  ``NNLO nucleon parton distributions from a light-cone quark model dressed 
with its virtual meson cloud",
  Phys.\ Rev.\ D {\bf 89} (2014) 034021 [arXiv:1309.5814 [hep-ph]]~.


  
  
\bibitem{Dirac}
  P.~A.~M.~Dirac,
  ``Forms of Relativistic Dynamics",
  Rev.\ Mod.\ Phys.\  {\bf 21} (1949) 392~.


\bibitem{Mekhfi:1988kj} 
  M.~Mekhfi and X.~Artru,
  ``Sudakov Suppression Of Color Correlations In Multiparton Scattering",
  Phys.\ Rev.\ D {\bf 37}, 2618 (1988)~.

\bibitem{spinor0}
  G.~P.~Lepage and S.~J.~Brodsky,
  ``Exclusive Processes in Perturbative Quantum Chromodynamics'',
  Phys.\ Rev.\ D {\bf 22}, 2157 (1980).


\bibitem{spinor} 
  M.~Diehl, T.~Feldmann, R.~Jakob and P.~Kroll,
  ``The Overlap representation of skewed quark and gluon distributions",
  Nucl.\ Phys.\ B {\bf 596}, 33 (2001)
  [Erratum-ibid.\ B {\bf 605}, 647 (2001)] [hep-ph/0009255]~.
  
   \bibitem{BT}
  B.~Bakamjian and L.~H.~Thomas,
  ``Relativistic particle dynamics. 2'',
  Phys.\ Rev.\  {\bf 92}, 1300 (1953)~.
  
  
\bibitem{giannin} 
  M.~Ferraris, M.~M.~Giannini, M.~Pizzo, E.~Santopinto and L.~Tiator,
  ``A Three body force model for the baryon spectrum",
  Phys.\ Lett.\ B {\bf 364}, 231 (1995)~.
  
\bibitem{kperp} 
  D.~Boer, S.~J.~Brodsky and D.~S.~Hwang,
  ``Initial state interactions in the unpolarized Drell-Yan process'',
  Phys.\ Rev.\ D {\bf 67}, 054003 (2003)·~.

  
\bibitem{pasqmu}
  B.~Pasquini and P.~Schweitzer,
  ``Pion TMDs in light-front constituent approach, and Boer-Mulders effect 
in the pion-induced Drell-Yan process",
  Phys.\ Rev.\ D {\bf 90} (2014) 014050~.

\bibitem{cluster}
  S.~J.~Brodsky and C.~R.~Ji,
  ``Factorization Property of the Deuteron'',
  Phys.\ Rev.\ D {\bf 33}, 2653 (1986).

\bibitem{trelest}
  M.~Strikman and D.~Treleani,
  ``Measuring double parton distributions in nucleons at proton nucleus
colliders'',
  Phys.\ Rev.\ Lett.\  {\bf 88}, 031801 (2002).

\bibitem{simona}
  S.~Salvini, D.~Treleani and G.~Calucci,
  ``Double Parton Scatterings in High-Energy Proton-Nucleus Collisions 
and Partonic Correlations'',
  Phys.\ Rev.\ D {\bf 89}, 016020 (2014).

  

\end{thebibliography}
\end{document}